\documentclass{article}
\usepackage{amsmath,amssymb,tabularx,bm,booktabs,authblk}
\usepackage{xcolor,slashed,epsfig,cite}
\usepackage{verbatim,slashed,ulem,geometry,setspace,algorithm}
\usepackage{algorithmic,mathrsfs,braket,multirow,dcolumn,subfigure}
\allowdisplaybreaks[4]
\usepackage{graphicx}
\usepackage[colorlinks=true,citecolor=blue,urlcolor=cyan,filecolor=magenta,linkcolor=red,frenchlinks=true,bookmarks]{hyperref}
\allowdisplaybreaks[4]
\renewcommand\sout{\bgroup \color{red} \ULdepth=-.5ex \ULset}
\newcommand{\nobracket}{}
\newcommand{\email}[1]{\thanks{\href{mailto:#1}{#1}}}

\usepackage{color}

\definecolor{mycolor}{rgb}{0.6,0.0,0.4}

\definecolor{mycolorg}{rgb}{0.0,0.6,0.2}

\begin{document}

\title{Transversity generalized parton distributions in spin-3/2 particles}

\author[a,b]{Dongyan Fu\email{fudongyan@ihep.ac.cn}}
\author[a,b]{Yubing Dong\email{dongyb@ihep.ac.cn}}
\author[c,d,e]{S. Kumano\email{shunzo.kumano@kek.jp}}

\affil[a]{Institute of High Energy Physics, Chinese Academy of Sciences, Beijing 100049, China}
\affil[b]{School of Physical Sciences, University of Chinese Academy of Sciences,
Beijing 101408, China}

\affil[c]{Department of Mathematics, Physics, and Computer Science, Faculty of Science,\newline
Japan Women's University, Mejirodai 2-8-1, Tokyo 112-8681, Japan}
\affil[d]{Theory Center, Institute of Particle and Nuclear Studies, KEK
\newline
Oho 1-1, Tsukuba, Ibaraki, 305-0801, Japan}
\affil[e]{
{Institute of Modern Physics, Chinese Academy of Sciences, Lanzhou 730000, China}
}

\maketitle
\begin{abstract}
The definitions of the quark and gluon transversity generalized parton distributions (GPDs)
in spin-3/2 particles are obtained in the light-cone gauge. It is found that they contain
16 independent components for each parton. Their even or odd property is found in terms of skewness variable, and the odd transversity GPDs vanish in the forward
limit. There are 16 amplitudes with helicity flip of quark or gluon. We conclude
that all the amplitudes, unpolarized, polarized, and transversity amplitudes,
have a common factor $F(\zeta)$ carrying all the complex part.
Here, the variable $\zeta$ is the helicity difference
$\zeta=(\lambda' -\lambda)-(\mu' - \mu)$ for the amplitude $\mathcal{A}_{\lambda' \mu',\lambda\mu}$.
This kinematical factor is associated with the transfer
of the orbital angular momentum in the quark- and gluon-spin-3/2 particle
scattering amplitudes.
We also derive another main physical
quantity, transversity distribution, from the transversity GPDs in the forward limit for the
spin-3/2 particles.
\end{abstract}

\section{Introduction}\label{sectionintroduction}

\quad\,\
To investigate the construction of hadrons from quarks and gluons degrees of freedom,
which are the elementary particles described by quantum chromodynamics (QCD), the notion
of generalized parton distributions (GPDs) was introduced~\cite{Muller:1994ses, Ji:1996ek,
Radyushkin:1996nd, Radyushkin:1996ru,Ji:1996nm, Radyushkin:1997ki, Ji:1998pc}.
The GPDs can provide a more detailed account of the internal structure of hadrons in terms of their
elementary constituents than the parton distributions
and form factors (FFs). Similar
to the theoretical constructs, one of the related experiments utilized to investigate hadron
structures is the deep inelastic scattering process, which indicates the parton distributions,
including the unpolarized, polarized, and transversity distribution functions at the leading
twist. We know that the GPDs depend on three variables: the longitudinal momentum fraction $x$,
the square of the transferred momentum $t$, and the skewness variable $\xi$ which is defined
by the transferred longitudinal momentum, rather than just one variable $x$ in the parton distribution functions (PDFs).
The reaction that involves GPDs is the off-forward Compton scattering,
which consists of at least one off-shell photon~\cite{Ji:1998pc,Diehl:2003ny}, including
the deeply virtual Compton scattering 
(DVCS) 
process~\cite{Radyushkin:1996nd,Ji:1996nm}.
Furthermore, the GPDs reduce to the PDFs in the forward limit.
The spacelike GPDs are measured by the DVCS and meson-production processes 
at the Thomas Jefferson National Accelerator Facility (JLab) and 
in the CERN-AMBER project. 
The timelike GPDs are investigated by the two-photon process
or the timelike Compton scattering at KEKB \cite{PhysRevD.93.032003,PhysRevD.97.014020}
and will be studied possibly at BESIII. 
In addition, there are future GPD projects 
at EICs \cite{ABDULKHALEK2022122447,Anderle2021},
by hadron reactions at 
the Japan Proton Accelerator Research Complex (J-PARC)
\cite{Kumano:2009he,Sawada:2016mao,Qiu:2022bpq,Qiu:2022pla},
and by neutrino reactions at Fermilab
\cite{PhysRevD.95.114029,Kumano:2022DJ,Chen-neutrino}.

Corresponding to transverse momentum independent parton distributions~\cite{Barone:2003fy},
there are three types of GPDs, i) unpolarized
GPD corresponding to the unpolarized parton
distribution function, the probability of finding a parton with the longitudinal momentum
fraction $x$ in the unpolarized hadron; ii) polarized GPD
corresponding to the polarized
distribution function, the number density of a parton with the longitudinal momentum fraction
$x$ and spin parallel minus spin antiparallel to that of the polarized hadron;
iii) transversity
GPD corresponding to the transversity distribution function
$h_1(x)$~\cite{Hoodbhoy:1998vm,Cosyn:2018rdm} (called $\Delta_T (x)$ in
Ref.~\cite{Barone:2003fy}, $\Delta(x)$ in Ref.~\cite{Jaffe:1989xy} and $\delta q(x)$ in
Ref.~\cite{Diehl:2001pm}), which denotes the number density of a parton with the
longitudinal momentum fraction $x$ and polarization parallel to that of the hadron
with transverse polarization minus the number density with antiparallel polarization.
The transversity distribution function $h_1(x)$, which can be obtained from transversity GPDs
in the forward limit as described above, has been studied fundamentally by
Ralston et al.~\cite{Ralston:1979ys}, Artru et al.~\cite{Artru:1989zv},
Jaffe et al.~\cite{Jaffe:1991kp,Jaffe:1991ra},
Cortes et al.~\cite{Cortes:1991ja}, Ji~\cite{Ji:1992ev},
and Barone et al.~\cite{Barone:2001sp}. Additionally, there is no logical reason to assume
that the transversity parton distribution is significantly smaller than the other two
distributions~\cite{Barone:2001sp, Scopetta:1997qg, Suzuki:1997vu, Ma:1997gy, Gamberg:1998vg, Cloet:2007em, Kumano:2019igu}.
For nearly two decades, the spin-1/2 and -1 quark transversity
distributions have been measured by some
experiments~\cite{COMPASS:2005csq,Belle:2008fdv,Moretti:2018zjm,COMPASS:2021bws} and the gluon
transversity is possibly measured by 
JLab~\cite{jlab-gluon-trans},
EIC (Electron-Ion Collider)~\cite{Ye:2016prn,Accardi:2012qut,ABDULKHALEK2022122447},
and EicC (Electron-Ion Collider in China) \cite{Anderle2021},
and Fermilab~\cite{Kumano:2019igu,Fermilab-spin,Keller-2022}.

Based on the parity and time reversal invariances, there are $2 (2 J+1)^2$ independent GPDs
for each parton (either quark flavor or gluon) in a spin-$J$ hadron at leading twist~\cite{Diehl:2003ny}.
These independent GPDs are discussed in
Sec.~3.5.4 of Ref.~\cite{Diehl:2003ny} and Ref.~\cite{Fu:2022bpf} for spin-3/2 hadrons,
and it is also found by using the parity invariance in Eq.~\eqref{paritytime} of this paper.

Half of the total GPDs consist of the unpolarized and polarized GPDs that do not involve the
helicity flip, while the remaining half is distributed to the transversity GPDs involving the
helicity flip, i.e., there are 16 independent transversity GPDs for the
spin-3/2 hadrons. The quark transversity GPDs are chiral-odd since the corresponding
operator flips the quark chirality, in contrast to the chiral-even unpolarized and polarized
GPDs. Moreover, the helicity flip distributions of the quark and gluon will mix in the
evolution of the transversity GPDs. Many studies on the definition of
GPDs~\cite{Ji:1998pc, Diehl:2003ny, Diehl:2001pm, Berger:2001zb, Cosyn:2018rdm} and the model
calculations~\cite{Fanelli:2016aqc, Diehl:2013xca, Sun:2017gtz} have been conducted for the
low spin ($J \leq$1) hadrons.

It should be stressed that the unpolarized and polarized GPDs of spin-3/2 particles have been
decomposed, defined, and calculated in our recent papers~\cite{Fu:2022bpf,Fu:2023dea}. The
priority of this work is to decompose and derive the definitions of the transversity GPDs
for the spin-3/2 system. We expect that this series studies of GPDs for spin-3/2 particles
give valuable references for possible future experiments at EIC, EicC, BESIII, KEK-B factory, and J-PARC.
Moreover, the study of the gluon
transversity distribution on $\Delta$ isobar is also necessary to understand the non-nucleonic
degrees of the hadrons in nuclei~\cite{Nzar:1992ax, Jaffe:1989xy, Sather:1990bq, Miller:2013hla}.
Experimentally, the timelike GPDs could be measured in principle
at BESIII and KEK-B \cite{PhysRevD.97.014020}
even for spin-3/2 particles like
$e^+e^- \to \Delta\bar{\Delta}$ or $\Omega\bar{\Omega}$.
Furthermore, there are recent interests on the transition GPDs for $N \to \Delta$ at JLab and J-PARC~\cite{Kumano:2009he,transition-GPDs},
so that spin-3/2 GPD experimental project could become a realistic one in future.

The main content is concentrated in Sec.~\ref{sectiontgpds}. Firstly, the definitions of
transversity GPDs and their corresponding amplitudes, along with their symmetry
properties, are explained and discussed. In the next two subsections, the specific components
of the parton transversity GPDs respectively for the quark and gluon, and the interpretations
of their symmetry properties are shown. The corresponding tensor form factors and tensor charges are obtained by the derived sum rules. Moreover, the amplitudes and
the relations connecting transversity GPDs with transversity parton distributions are
derived and elucidated.  To summarize our work, we provide a short discussion in
Sec.~\ref{sectionsummary}.

\section{Transversity GPDs of spin-3/2 particles}\label{sectiontgpds}

\subsection{Conventions, correlators and symmetry properties}

\begin{figure}[h]
    \centering
    \includegraphics[height=2cm]{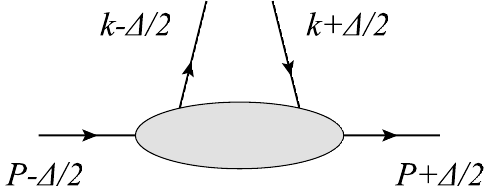} \quad
    \includegraphics[height=2cm]{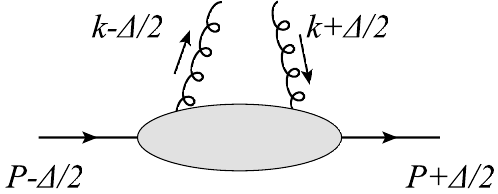}
    \caption{\small{Diagrams describing the quark (left) and gluon (right) GPDs.}}
    \label{figure1}
\end{figure}

\quad\,\
The GPDs are defined through matrix elements of the non-local parton operators. Following the
conventions of Ji~\cite{Ji:1998pc,Diehl:2003ny}, the leading twist GPDs can be
defined by the matrix element 
\begin{equation}\label{correlationfunction}
    \left.\int \frac{\text{d} z^-}{2 \pi} e^{i x \left( P \cdot z \right)} \left\langle p',
    \lambda' \left| \mathcal{O} \right|p,\lambda \right\rangle \right| _{z^+=0, \bm{z}_\perp=0},
\end{equation}
where $\mathcal{O}$ is the non-local operator at a light-like separation of the corresponding GPDs.
The $p$ ($p'$) and $\lambda$ ($\lambda'$) respectively denote the momentum and helicity
of the initial (final) state. In this work, the light-cone coordinate is employed and
any four-vector $v$ can be rewritten as $v=(v^+,v^-,\bm{v}_\perp)$, where $v^\pm = v^0 \pm v^3$
and $\bm{v}_\perp=(v^1, v^2)$. The scalar product of any two four-vectors is
$u \cdot v=\frac{1}{2} u^+ v^- +\frac{1}{2} u^- v^+ -\bm{u}_\perp \cdot \bm{v}_\perp$.
Moreover, the light-cone vector $n=(0,2,\bm{0}_\perp)$ is needed and $n^2=0$.
In addition, we use the same kinematical variables with our previous work~\cite{Fu:2022bpf},
\begin{equation}
    P=\frac{p'+p}{2}, \quad \Delta=p'-p, \quad t=\Delta^2, \quad \xi =
    - \frac{\Delta^+}{2 P^+} \,(|\xi| \leq 1),
    \quad x=\frac{k^+}{P^+}\,(-1 \leq x \leq 1),
\end{equation}
where $k-\Delta/2$ (or $k+\Delta/2$) is the initial (or final) parton momentum as
represented in Fig.~\ref{figure1}. Note that here we use $\Delta$ instead of $q$ in
Ref.~\cite{Fu:2022bpf} to stand for the relative momentum.
There are some other conventions in the following, like
$a^{[ \mu \nobracket} b^{\nobracket \nu ]}=a^\mu b^\nu - a^\nu b^\mu$,
$a^{\{ \mu \nobracket} b^{\nobracket \nu \}}=a^\mu b^\nu + a^\nu b^\mu$,
$\sigma^{n i}=\sigma^{\rho i} n_{\rho}$,
$\sigma^{\mu\nu} = (i/2) [\gamma^\mu,\gamma^\nu]$,
$\epsilon^{n i \rho \delta}=\epsilon^{\mu i \rho \delta} n_\mu$, and $\epsilon_{0123}=1$.

In Ref.~\cite{Fu:2022bpf}, the unpolarized (polarized) quark or gluon GPDs are defined and
decomposed for a spin-3/2 system when we choose the non-local operator being
$\mathcal{O}^q=\bar{\psi} \slashed{n} \psi$ ($\bar{\psi} \slashed{n} \gamma^5 \psi$)
or $\mathcal{O}^g=G^{n \mu} G_\mu ^{\ \, n}$ ($G^{n \mu} \Tilde{G}_\mu ^{\ \, n}$)
where $G^{n\mu} = G^{\rho \mu} n_\rho$
in Eq.~\eqref{correlationfunction}. In the present work, $\psi(x)$ and $G^{\mu \nu}(x)$ are
employed to respectively denote the quark field and gluon field strength tensor, and the
corresponding dual field strength tensor is defined as
$\Tilde{G}^{\rho \sigma} (x)=\frac{1}{2} \epsilon^{\rho \sigma \mu \nu} G_{\mu \nu}(x)$.
There is a Wilson line $W \left[ -\frac{1}{2} z^-,\frac{1}{2} z^- \right]$ along a
light-like path between the two fields at positions of $-\frac{1}{2}z^-$ and $\frac{1}{2}z^-$
in the non-local operator $\mathcal{O}$. The Wilson line is defined as~\cite{Diehl:2003ny}
\begin{equation}
    W \left[ a, b \right] = \mathcal{P}\, \text{exp}
    \left[ i g \int^a_b d x^- A^+(x^- n_-)\right],
\end{equation}
where $\mathcal{P}$ stands for the path-ordering from $b$ to $a$ and $A(x)$
represents the gluon field. In this work, the light-cone gauge $A^+=0$ for the gluon field is
considered and then
$W \left[ -\frac{1}{2} z^-,\frac{1}{2} z^- \right]=1$,
so that the Wilson line does not appear in the operators.

Equation~\eqref{correlationfunction} with non-local operators of
$\mathcal{O}^q=\bar{\psi}(-\frac{1}{2}z) i \sigma^{n i} \psi(\frac{1}{2}z)$ and
$\mathcal{O}^g=G^{n i}(-\frac{1}{2}z) G^{j n}(\frac{1}{2}z)$ defines the quark and gluon
transversity GPDs as
\begin{equation}\label{quarktransversitygpds}
    T^{q i}_{\lambda' \lambda}=\frac{1}{2} \int \frac{\text{d} z^-}{2 \pi}
    e^{i x \left( P \cdot z \right)} \left.
    \left\langle p', \lambda' \left| \bar{\psi} (-\frac{1}{2}z) i \sigma^{n i} \psi
    (\frac{1}{2}z) \right|p,
    \lambda \right\rangle \right| _{z^+=0, \bm{z}_\perp=0},
\end{equation}
and
\begin{equation}\label{gluontransversitygpds}
    T^{g i j}_{\lambda' \lambda}=\frac{1}{2 P^+} \int \frac{\text{d} z^-}{2 \pi}
    e^{i x \left( P \cdot z \right)}
    \left.\left\langle p', \lambda' \left| \text{Tr}\hat{\bm{S}} \left[G^{n i}
    (-\frac{1}{2}z) G^{j n}
    (\frac{1}{2}z)\right] \right|p,\lambda \right\rangle \right| _{z^+=0, \bm{z}_\perp=0},
\end{equation}
where $i$ and $j$ are the transverse indices, and the operator $\hat{\bm{S}}$ in
Eq.~\eqref{gluontransversitygpds} stands for the removal of the trace and the
symmetrization between $i$ and $j$.
As the transversity GPDs have different definitions from the helicity non-flip GPDs, the quark and gluon transversity GPDs have the different decompositions since
gluon has the more index $j$.

One can also define the corresponding quark and gluon amplitudes~\cite{Diehl:2003ny}
as
\begin{equation}\label{helicityamplitude}
    \begin{split}
        \mathcal{A}^{q}_{\lambda'\mu', \lambda\mu}&=\int \frac{\text{d} z^-}{2 \pi}
        e^{i x \left( P \cdot z \right)}
        \left.\left\langle p', \lambda' \left| \mathcal{O}^q_{\mu' \mu}(z) \right|p,
        \lambda \right\rangle
        \right| _{z^+=0, \bm{z}_\perp=0},\\
        \mathcal{A}^{g}_{\lambda'\mu', \lambda\mu}&=\frac{1}{P^+}\int \frac{\text{d} z^-}{2 \pi}
        e^{i x \left( P \cdot z \right)} \left.\left\langle p', \lambda' \left|
        \mathcal{O}^g_{\mu' \mu}(z)
        \right|p,\lambda \right\rangle \right| _{z^+=0, \bm{z}_\perp=0},
    \end{split}
\end{equation}
where the label $\mu$ ($\mu'$) represents the emitted (re-absorbed) parton helicity.
The amplitudes satisfy the constraints of~\cite{Diehl:2003ny,Cosyn:2018rdm}
\begin{equation}\label{paritytime}
\begin{split}
    &\mathcal{A}^{q/g}_{-\lambda'-\mu', -\lambda-\mu}(P,\Delta,n)=(-1)^{(\lambda'-\lambda)-(\mu'-\mu)}
    \mathcal{A}^{q/g *}_{\lambda'\mu', \lambda\mu}(P,\Delta,n)\\
    &\mathcal{A}^{q/g}_{-\lambda' -\mu', -\lambda, -\mu}(P,\Delta,n)=(-1)^{(\lambda'-\lambda)-(\mu'-\mu)} \mathcal{A}^{q/g}_{\lambda \mu, \lambda' \mu'}(P,-\Delta,n)
\end{split}
\end{equation}
from hermiticity, parity and time reversal transformations.
Table~\ref{tableoperator} lists all the operators carrying the parton helicity at leading twist.
\begin{table}[h]
    \centering
	\begin{tabular}{ c  c  c  c }
		\toprule
		$\mathcal{O}^q_{\mu' \mu}$  & $\mathcal{O}^g_{\mu' \mu}$  & $\mu'$  & $\mu$\\
		\midrule
		$\frac{1}{4} \bar{\psi} \gamma^+(1+\gamma^5)\psi$    & $\frac{1}{2}
		\left[ G^{+\rho}G^{\,+}_\rho-i G^{+\rho}\Tilde{G}^{\,+}_\rho \right]$    & $+$ & $+$\\
		\rule{0pt}{13pt}
		$\frac{1}{4} \bar{\psi} \gamma^+(1-\gamma^5)\psi$
		& $\frac{1}{2}\left[ G^{+\rho}G^{\,+}_\rho
		+i G^{+\rho}\Tilde{G}^{\,+}_\rho \right]$    & $-$ & $-$\\
		\rule{0pt}{13pt}
		$-\frac{1}{4} \bar{\psi}(i \sigma^{+1}-i i\sigma^{+2})\psi$
		& $\frac{1}{2}\left[ G^{+1}G^{1+}-G^{+2}G^{2+}
		-i\left( G^{+1}G^{2+}+G^{+2}G^{1+} \right) \right]$    & $-$ & $+$\\
		\rule{0pt}{13pt}
		$\frac{1}{4} \bar{\psi}(i \sigma^{+1}+i i\sigma^{+2})\psi$
		& $\frac{1}{2}\left[ G^{+1}G^{1+}-G^{+2}G^{2+}
		+i\left( G^{+1}G^{2+}+G^{+2}G^{1+} \right) \right]$    & $+$ & $-$\\
		\bottomrule
	\end{tabular}
	\caption{\small{Leading twist operators $\mathcal{O}^q_{\mu' \mu}$ and
	$\mathcal{O}^g_{\mu' \mu}$.}}
	\label{tableoperator}
\end{table}
Notice that the helicity flip operators on the quark have another forms,
\begin{equation}
    \begin{split}
        \mathcal{O}^q_{- +} &= -\frac{1}{4} \bar{\psi}(i \sigma^{+1}-i
        i\sigma^{+2})\psi = -\frac{1}{4}
        \bar{\psi}i \sigma^{+1}(1+\gamma^5)\psi,\\ \mathcal{O}^q_{+ -} &
        = \frac{1}{4} \bar{\psi}(i \sigma^{+1}
        +i i\sigma^{+2})\psi = \frac{1}{4} \bar{\psi}i \sigma^{+1}(1-\gamma^5)\psi,
    \end{split}
\end{equation}
because of
$i \sigma^{+1} \gamma^5 = -i i \sigma^{+2}$.
Amplitudes~\eqref{helicityamplitude} with $\mu' = \mu$ contain 16 non-flip amplitudes in the spin-3/2 system, which have been explicitly given in
Ref.~\cite{Fu:2022bpf}. Analogously, the parity invariance restricts 16 independent quark
or gluon transversity GPDs for the spin-3/2 particles when $\mu' \neq \mu$. Note that
the helicity label $\pm$ on the partons of the quark stands for $\pm \frac{1}{2}$ and
those of the gluon for $\pm 1$ according to the corresponding spin. More symmetry
properties, such as hermiticity, light-front parity, and light-front time reversal,
are also satisfied~\cite{Cosyn:2018rdm}, and they do not provide any further constraints
on the number of GPDs.

\subsection{Quark transversity GPDs for a spin-3/2 particle}\label{subsectionquark}

\quad\,\ 
The quark transversity GPDs are defined by the matrix elements of
transverse non-local quark-quark correlator in Eq.~\eqref{quarktransversitygpds} as
\begin{equation}\label{quarkde}
    T^{q i}_{\lambda' \lambda} = -\bar{u}_{\alpha'} (p',\lambda')
    \mathcal{H}^{q T, i, \alpha' \alpha}(x,\xi,t) u_\alpha(p,\lambda),
\end{equation}
where $u_\alpha(p,\lambda)$ is the spin-3/2 field Rarita-Schwinger spinor, shown
in \ref{appendixconventions}, normalized to
$\bar{u}_{\alpha} (p,\lambda') u^\alpha(p,\lambda)=-2 M \delta_{\lambda' \lambda}$.
The tensor function $\mathcal{H}^{q T, i, \alpha' \alpha}(x,\xi,t)$ can be decomposed to
16 terms corresponding to only 16 independent tensor structures. Similar to our previous
analysis~\cite{Fu:2022bpf}, there are two ways to obtain the sufficiently tensor structures:
1) write down all the possible structures;
2) utilize the direct product between the
transvesity spin-1/2 and unpolarized spin-1 structures or between the transversity
spin-1 and unpolarized spin-1/2.
We can then reduce the redundant tensor structures using some on-shell identities (seen also
\ref{appendixidentities}). Consequently, the tensor function of
$\mathcal{H}^{q T, i, \alpha' \alpha}$ is decomposed at twist 2 by imposing Hermiticity, parity invariance,
and time-reversal invariance as
\begin{equation}\label{quarkdecompositions}
\begin{split}
    \mathcal{H}^{q T, i, \alpha' \alpha}= & H^{q T}_1 \frac{i \sigma^{n i}}
    {\left( P \cdot n \right)} g^{\alpha' \alpha}
    + H^{q T}_2 \frac{n^{[ \alpha' \nobracket} g^{\nobracket \alpha ] i}}
    {\left( P \cdot n \right)}
    + H^{q T}_3 \frac{\left( \slashed{n} P^i - P \cdot n \, \gamma^i \right)}
    {M \left( P \cdot n \right)} g^{\alpha' \alpha}
    + H^{q T}_4 \frac{\left( \slashed{n} P^i - P \cdot n \, \gamma^i \right)}
    {M^3 \left( P \cdot n \right)} P^{\alpha'} P^\alpha \\
    & + H^{q T}_5 \frac{\left( \slashed{n} \Delta^i - \Delta \cdot n \,
    \gamma^i \right)}{M \left( P \cdot n \right)} g^{\alpha' \alpha}
    + H^{q T}_6 \frac{\left( \slashed{n} \Delta^i - \Delta \cdot n \, \gamma^i \right)}{M^3
    \left( P \cdot n \right)} P^{\alpha'} P^\alpha \\
	& + H^{q T}_7 \frac{\left(\Delta^i + 2 \xi P^i \right)}{M^2}g^{\alpha' \alpha}
	+ H^{q T}_8 \frac{\left(\Delta^i + 2 \xi P^i \right)}{M^4}P^{\alpha'} P^\alpha\\
	& + H^{q T}_{9} \frac{\left( \Delta \cdot n \, n^{\{ \alpha' \nobracket}
	g^{\nobracket \alpha \} i} -2 n^{\alpha'}
	n^\alpha \Delta^i \right)}{\left( P \cdot n \right)^2}
	+ H^{q T}_{10} \frac{\left( P \cdot n \, n^{\{ \alpha' \nobracket}
	g^{\nobracket \alpha \} i}
	-2 n^{\alpha'} n^\alpha P^i \right)}{\left( P \cdot n \right)^2}\\
	& + H^{q T}_{11} \frac{\left( \Delta \cdot n \, P^{[ \alpha' \nobracket}
	g^{\nobracket \alpha ] i}- P^{[ \alpha' \nobracket} n^{\nobracket \alpha ]}
	\Delta^i \right)}{M^2 \left( P \cdot n \right)}
	+ H^{q T}_{12} \frac{\left( P \cdot n \, P^{[ \alpha' \nobracket} g^{\nobracket \alpha ] i}
	- P^{[ \alpha' \nobracket} n^{\nobracket \alpha ]} P^i \right)}
	{M^2 \left( P \cdot n \right)}\\
	& + H^{q T}_{13} \frac{M \slashed{n} \left( \Delta \cdot n n^{\{ \alpha' \nobracket}
	g^{\nobracket \alpha \} i} - 2 n^{\alpha'}
	n^\alpha \Delta^i \right)}{\left( P \cdot n \right)^3}
	+ H^{q T}_{14} \frac{M \slashed{n} \left( P \cdot n n^{\{ \alpha' \nobracket}
	g^{\nobracket \alpha \} i} - 2 n^{\alpha'} n^\alpha P^i \right)}
	{\left( P \cdot n \right)^3}\\
    & + H^{q T}_{15} \frac{\slashed{n} \left( \Delta \cdot n \, P^{[ \alpha' \nobracket}
    g^{\nobracket \alpha ] i}- P^{[ \alpha' \nobracket} n^{\nobracket \alpha ]}
    \Delta^i \right)}{M \left( P \cdot n \right)^2}
    + H^{q T}_{16} \frac{\slashed{n} \left( P \cdot n \, P^{[ \alpha' \nobracket}
    g^{\nobracket \alpha ] i}- P^{[ \alpha' \nobracket} n^{\nobracket \alpha ]}
    P^i \right)}{M \left( P \cdot n \right)^2},
\end{split}
\end{equation}
where the variables $x$, $\xi$, $t$ in
the quark transversity GPDs $H^{q T}_i$ are omitted,
and $M$ is the mass of the spin-3/2 particle.

There is another equivalent definition of quark transversity GPDs using the matrix element of
$i \sigma^{n i} \gamma^5$ instead of $i \sigma^{n i}$ in Eq.~\eqref{quarktransversitygpds}.
Due to the relation~\cite{Diehl:2001pm}
\begin{equation}
    i \sigma^{\alpha \beta} \gamma^5 = -\frac{i}{2}
    \epsilon^{\alpha \beta \rho \delta} i \sigma_{\rho \delta},
\end{equation}
thus the equivalent definition is
\begin{equation}\label{quarktransversitygpdsgamma5}
    \frac{1}{2} \int \frac{\text{d} z^-}{2 \pi}
    e^{i x \left( P \cdot z \right)} \left.\left\langle p', \lambda' \left| \bar{\psi}
    (-\frac{1}{2}z) i \sigma^{n i} \gamma^5 \psi (\frac{1}{2}z) \right|p,\lambda \right\rangle
    \right| _{z^+=0, \bm{z}_\perp=0}=-\bar{u}_{\alpha'} (p',\lambda')
    \tilde{\mathcal{H}}^{q T, i, \alpha' \alpha}(x,\xi,t) u_\alpha(p,\lambda),
\end{equation}
where
\begin{equation}\label{quarkdecompositionsgamma5}
\begin{split}
    \tilde{\mathcal{H}}^{q T, i, \alpha' \alpha} =& H^{q T}_1
    \frac{i \sigma^{n i} \gamma^5}{\left( P \cdot n \right)} g^{\alpha' \alpha}
    +  H^{q T}_2 \frac{i \epsilon^{n i \alpha \alpha'} }{\left( P \cdot n \right)}
    + H^{q T}_3 \frac{i \epsilon^{n i P \delta}  \gamma_\delta  g^{\alpha' \alpha}}
    {M \left( P \cdot n \right)}
    + H^{q T}_4 \frac{i \epsilon^{n i P \delta} \gamma_\delta  P^{\alpha'} P^\alpha}{M^3
    \left( P \cdot n \right)} \\
    & + H^{q T}_5 \frac{i \epsilon^{n i \Delta \delta} \gamma_\delta g^{\alpha' \alpha} }
    {M \left( P \cdot n \right)}
    + H^{q T}_6 \frac{i \epsilon^{n i \Delta \delta} \gamma_\delta P^{\alpha'} P^\alpha }
    {M^3  \left( P \cdot n \right)}
	+ H^{q T}_7 \frac{i \epsilon^{n i \Delta P} g^{\alpha' \alpha}}
	{M^2 \left( P \cdot n \right)}
	+ H^{q T}_8 \frac{i \epsilon^{n i \Delta P} P^{\alpha'} P^\alpha}
	{M^4 \left( P \cdot n \right)}\\
	& + H^{q T}_{9} \frac{i \epsilon^{n i \{ \alpha' \nobracket \Delta}
	n^{\nobracket \alpha \} } }{\left( P \cdot n \right)^2}
	+ H^{q T}_{10} \frac{i \epsilon^{n i \{ \alpha' \nobracket P}
	n^{\nobracket \alpha \} }}{\left( P \cdot n \right)^2}
	+ H^{q T}_{11} \frac{i \epsilon^{n i \Delta [ \alpha' \nobracket} P^{\alpha ]}}
	{M^2 \left( P \cdot n \right)}
	+ H^{q T}_{12} \frac{i \epsilon^{n i P [ \alpha' \nobracket} P^{\alpha ]}}
	{M^2 \left( P \cdot n \right)}\\
	& + H^{q T}_{13} \frac{M \slashed{n} \, i \epsilon^{n i \{ \alpha' \nobracket \Delta}
	n^{\nobracket \alpha \} }}{\left( P \cdot n \right)^3}
	+ H^{q T}_{14} \frac{M \slashed{n} \, i \epsilon^{n i \{ \alpha' \nobracket P}
	n^{\nobracket \alpha \} }}{\left( P \cdot n \right)^3}
    + H^{q T}_{15} \frac{\slashed{n} \, i \epsilon^{n i \Delta [ \alpha' \nobracket}
    P^{\alpha ]}}{M \left( P \cdot n \right)^2}
    + H^{q T}_{16} \frac{\slashed{n} \, i \epsilon^{n i P [ \alpha' \nobracket}
    P^{\alpha ]}}{M \left( P \cdot n \right)^2},
\end{split}
\end{equation}
which is given by
$ \tilde{\mathcal{H}}^{q T, i', \alpha' \alpha}
 = i \epsilon^{0 i' i 3} \mathcal{H}^{q T, i, \alpha' \alpha}$.
The quark transversity GPDs $H^{q T}_i$ in Eq.~\eqref{quarkdecompositionsgamma5}
is the same as the corresponding ones in Eq.~\eqref{quarkdecompositions}.

The relations of the amplitudes in Eq.~\eqref{paritytime} determines the even or odd property with respect to the variable
of skewness $\xi$ as
\begin{equation}
    \begin{split}
        H^{q T}_i(x,\xi,t) & =H^{q T}_i(x,-\xi,t) \quad \text{with}
        \quad i=1,2,5 \sim 9,12,13,16,\\
        H^{q T}_j(x,\xi,t) & =-H^{q T}_j(x,-\xi,t) \quad \text{with} \quad j=3,4,10,11,14,15,\\
    \end{split}
\end{equation}
where $H^{q T}_{3,4,10,11,14,15}$ are odd and others are even in $\xi$.
Clearly, all the odd GPDs go to zero when $\xi=0$.

It might be possibly insignificant but necessary to show the corresponding tensor FFs. In
comparison with the unpolarized and polarized GPDs, one can consider the local
quark-quark operator corresponding to quark
transversity GPDs~\eqref{quarktransversitygpds} as
\begin{equation}\label{formfactors}
    T^{\mu \nu}=\left\langle p', \lambda' \left| \bar{\psi} (0)
    i \sigma^{\mu \nu} \psi (0) \right|p,\lambda \right\rangle=-2\,
    \bar{u}_{\alpha'} (p',\lambda') \mathcal{F}_q^{\mu \nu,\alpha' \alpha} u_\alpha(p,\lambda),
\end{equation}
which give the $\xi$-independant tensor FFs.
Since the first moments of GPDs in Eq.~\eqref{quarkdecompositions} give the tensor FFs
in Eq.~\eqref{formfactors} by sum rules, the tensor structures corresponding to
tensor FFs are contained in Eq.~\eqref{quarkdecompositions}. In addition, the multi-$n$ terms,
like $\Delta \cdot n \, n^{\{ \alpha' \nobracket} g^{\nobracket \alpha' \} i}$, $n^{\alpha'}
n^\alpha$, $\Delta \cdot n \, \slashed{n} \, n^{\{ \alpha' \nobracket}
g^{\nobracket \alpha' \} i}$
and so on, should not exist in the tensor structures corresponding to tensor FFs because
there is only a position, $\mu$ or $\nu$, permitted to place $n$.

Considering the transversity GPDs~\eqref{quarkdecompositions} and the terms with non-vanishing
integration, we get the decomposition of Eq.~\eqref{formfactors} in terms of the tensor FFs as
\begin{equation}
    \begin{split}
        \mathcal{F}_q^{\mu \nu,\alpha' \alpha}= & g^{\alpha' \alpha}\left(G^{q T}_1(t)
        i \sigma^{\mu \nu} + G^{q T}_5(t) \frac{\gamma^{[ \mu \nobracket}
        \Delta^{\nobracket \nu ]}}{M}
        + G^{q T}_7(t) \frac{P^{[ \mu \nobracket} \Delta^{\nobracket \nu ]}}{M^2} \right) \\
        & + \frac{P^{\alpha'} P^\alpha}{M^2}\left( G^{q T}_6(t)
        \frac{\gamma^{[ \mu \nobracket} \Delta^{\nobracket \nu ]}}{M} + G^{q T}_8(t)
        \frac{P^{[ \mu \nobracket} \Delta^{\nobracket \nu ]}}{M^2} \right)
        + G^{q T}_{2}(t) g^{\mu [ \alpha' \nobracket} g^{\nobracket \alpha ] \nu}
        + G^{q T}_{12}(t) \frac{P^{[\alpha' \nobracket} g^{\nobracket \alpha ]
        [\nu \nobracket} P^{\nobracket \mu ]}}{M^2},
    \end{split}
\end{equation}
and the tensor FFs $G^{q T}_i(t)$ are related with the transversity GPDs by the sum rules
\begin{equation}\label{quarksumrules}
\begin{split}
    \int ^1_{-1} d x \, H^{q T}_i(x,\xi,t)&=G^{q T}_i (t) \quad \text{with}
    \quad i=1,2,5 \sim 8,12,\\
    \int ^1_{-1} d x \, H^{q T}_j(x,\xi,t)&=0 \quad \text{with}
    \quad j=3,4,9,10,11,13 \sim 16.
\end{split}
\end{equation}
There are three~\cite{Diehl:2001pm}, five~\cite{Cosyn:2018rdm} and seven tensor FFs for the spin-1/2, -1 and -3/2 system, respectively.
Since there is the relation, $\bar{u}_{\alpha'} (p,\lambda') \left( g^{\alpha' \alpha} i \sigma^{\mu \nu} + g^{\mu [ \alpha' \nobracket} g^{\nobracket \alpha ] \nu} \right) u_\alpha(p,\lambda) = 0$, in the forward limit, which is proved in~\ref{appendixidentities}, there is only a nonzero form factor $G^{q T}_1(0)-G^{q T}_2(0)$ describing the quark tensor charge.
Beyond the Standard Model, the tensor charge is explained as the number of the electric dipole moment carried by the corresponding quark~\cite{Pospelov:2005pr,Wang:2018kto}.

In addition, the quark amplitudes with helicity flip can be derived according to their definitions in
Eq.~\eqref{helicityamplitude}. We use the same notations as in our previous
work~\cite{Fu:2022bpf}
\begin{equation}\label{helicityamplitudeconventions}
\begin{split}
    &| \bm{p}_{\bot} |e^{\pm i \phi} \equiv p^x \pm i p^y, \quad
    | \bm{p}'_{\bot} |e^{\pm i \phi'} \equiv p'^x \pm i p'^y,\\
    & C= \sqrt{\frac{1 - \xi}{1 + \xi}} \frac{| \bm{p}_{\bot} |}{M} e^{- i \phi} -
    \sqrt{\frac{1 + \xi}{1 - \xi}} \frac{| \bm{p}'_{\bot} |}{M} e^{- i \phi'}
    = - \frac{(\Delta+2\xi P)^x-i (\Delta+2\xi P)^y}{M \sqrt{1-\xi^2}},\\
    & D = - \frac{t}{4 M^2} - \frac{\xi^2}{1 - \xi^2}.
\end{split}
\end{equation}
The notation $C$ represents the transfer of the transverse momentum between the initial and
final hadron states, which induces that the hadron obtains the transfer of the orbital angular
momentum that violates the helicity conservation, i.e. $\lambda'-\lambda=\mu'-\mu$ for
$\mathcal{A}_{\lambda'\mu', \lambda\mu}$. In Eq.~\eqref{helicityamplitudeconventions}, the transversity indices $x$ and $y$ instead of $1$ and $2$ are used to avoid ambiguity.
Moreover, $C$ will be real when the azimuthal angle of the four-vector
$\Delta+2 \xi P$~\cite{Diehl:2003ny,Cosyn:2018rdm} is zero, i.e. $(1-\xi) p^y =(1+\xi) p'^y$,
and be zero in the forward limit.

Incorporating the amplitudes with helicity non-flip in Ref.~\cite{Fu:2022bpf}, we find that any
amplitude $\mathcal{A}_{\lambda' \mu', \lambda \mu}$ has a common factor
\begin{equation}\label{commonfactor}
    F(\zeta)=C^\zeta \theta(\zeta)+C^{\ast -\zeta} \theta(-\zeta) \quad \text{with} \quad \zeta=(\lambda'-\lambda)-(\mu'-\mu),
\end{equation}
where $C^\zeta$ represents the $\zeta$ powers of $C$, $\theta(\zeta)$ is the Heaviside step function with $\theta(0)=\frac{1}{2}$ and $F(0)=1$.
We conclude that $F(\zeta)$ has some properties, such as $F(\zeta \neq 0)=0$ in the forward limit
and $F^\ast(\zeta) = F(-\zeta)$. Therefore, the amplitudes are real if the factor $F(\zeta)$
is excluded and we can define the real amplitude as
\begin{equation}\label{helicityamplitudereal}
    \mathcal{A}'_{\lambda'\mu', \lambda\mu}= \frac{\mathcal{A}_{\lambda'\mu', \lambda\mu}}
    {F(\lambda'-\lambda-\mu'+\mu)},
\end{equation}
and the relations in Eq.~\eqref{paritytime} becomes the real form,
\begin{equation}\label{parityinvariancereal}
    \mathcal{A}'_{-\lambda'-\mu', -\lambda-\mu}=(-1)^{(\lambda'-\lambda)-(\mu'-\mu)}
    \mathcal{A}'_{\lambda'\mu', \lambda\mu}.
\end{equation}

For the quark,
the factors $\mu'$ and $\mu$ are
$\mu' \, (\mu)=\pm \frac{1}{2}$ due to the quark spin.
We believe that the low spin particles~\cite{Diehl:2001pm,Cosyn:2018rdm} have
the analogous character \footnote{There might be a typo in (C15) of Ref.~\cite{Cosyn:2018rdm}
according to this property.}. The specific forms of the 16 quark amplitudes are
\begin{equation}
  \begin{split}
    \mathcal{A}'^{\, q}_{(3 / 2) -, (3 / 2) +} =
    &(1 + \xi) \left[ \left( \frac{1}{2} H^{q T}_3 - H^{q T}_5 \right) + \frac{| C
    |^2}{8} \left( \frac{1}{2} H^{q T}_4 - H^{q T}_6 \right) - \left( \xi H^{q T}_{15} -
    \frac{1}{2} H^{q T}_{16} \right) \right] - \left( H^{q T}_7 + \frac{| C |^2}{8}
    H^{q T}_8 \right)\\
    &- \frac{1}{1 - \xi} \left( \xi H^{q T}_{11} - \frac{1}{2} H^{q T}_{12} \right),
  \end{split}
\end{equation}
\begin{equation}
  \begin{split}
    \mathcal{A}'^{\, q}_{(3 / 2) -, (1 / 2) +} =
    &- \frac{(1 + \xi)}{2 \sqrt{3 (1 - \xi^2)}} [(1 + \xi) (H^{q T}_3 - 2 H^{q T}_5) -
    (3 - \xi) H^{q T}_7 + (1 - \xi) H^{q T}_{11} + H^{q T}_{12}]\\
    &+ \frac{\sqrt{1 - \xi^2}}{2 \sqrt{3}} \left[ \frac{| C |^2}{8} H^{q T}_8 -
    (1 + \xi) \left( H^{q T}_{15} + \frac{1}{2} H^{q T}_{16} \right) \right]\\
    &- \frac{ [D (1 - \xi^2) + \xi]}{2 \sqrt{3 (1 - \xi^2)}} \left[ \frac{1 +
    \xi}{1 - \xi} \left( \frac{1}{2} H^{q T}_4 - H^{q T}_6 \right) - \frac{1}{1 - \xi}
    H^{q T}_8 \right],
  \end{split}
\end{equation}
\begin{equation}
  \begin{split}
    \mathcal{A}'^{\, q}_{(3 / 2) -, (- 1 / 2) +} =
    &- \frac{(1 + \xi)}{8 \sqrt{3}} \left( \frac{1}{2} H^{q T}_4 - H^{q T}_6 + 4 H^{q T}_7
    - 2 H^{q T}_{11} - H^{q T}_{12} \right) + \frac{1}{8 \sqrt{3}} H^{q T}_8 - \frac{[D (1
    - \xi^2) + \xi]}{4 \sqrt{3} (1 - \xi)} H^{q T}_8,
  \end{split}
\end{equation}
\begin{equation}
  \begin{split}
    \mathcal{A}'^{\, q}_{(3 / 2) -, (- 3 / 2) +}= - \frac{\sqrt{1 -
    \xi^2}}{16} H^{q T}_8,
  \end{split}
\end{equation}
\begin{equation}
  \begin{split}
    \mathcal{A}'^{\, q}_{(1 / 2) -, (3 / 2) +} =
    &- \frac{2 \sqrt{1 - \xi^2}}{\sqrt{3}} \left[ H^{q T}_1 - \frac{| C |^2}{2}
    \left( \frac{1}{2} H^{q T}_3 - H^{q T}_5 \right) + \frac{| C |^4}{32} H^{q T}_8
    \right]\\
    &- \frac{4 \sqrt{1 - \xi^2} (1 - \xi)}{\sqrt{3}} \left[ \left( \xi H^{q T}_{13}
    - \frac{1}{2} H^{q T}_{14} \right) + \frac{| C |^2}{8} \left. \left(
    H^{q T}_{15} - \frac{1}{2} H^{q T}_{16} \right. \right) \right]\\
    &+ \frac{2 (1 - \xi)}{\sqrt{3 (1 - \xi^2)}} \left[ H^{q T}_2 - 2 \left( \xi
    H^{q T}_9 - \frac{1}{2} H^{q T}_{10} \right) - \frac{| C |^2}{4} (3 + \xi) H^{q T}_7
    \right]\\
    &- \frac{2}{\sqrt{3 (1 - \xi^2)}} \left[ 2 \xi \left( \frac{1}{2} H^{q T}_3 -
    \xi H^{q T}_5 \right) + \frac{| C |^2}{4} \left( \xi \left( \frac{1}{2}
    H^{q T}_4 - \xi H^{q T}_6 \right) + (1 + \xi^2) H^{q T}_{11} - H^{q T}_{12} \right)
    \right]\\
    &- \frac{4 [D (1 - \xi^2) - \xi]}{\sqrt{3 (1 - \xi^2)}} \left[ \frac{1}{1 -
    \xi^2} \left( \xi H^{q T}_{11} - \frac{1}{2} H^{q T}_{12} \right) + \left( \xi
    H^{q T}_{15} - \frac{1}{2} H^{q T}_{16} \right) - \frac{| C |^2}{8} \left(
    \frac{1}{2} H^{q T}_4 - H^{q T}_6 - \frac{1}{1 + \xi} H^{q T}_8 \right) \right],
  \end{split}
\end{equation}
\begin{equation}
  \begin{split}
    \mathcal{A}'^{\, q}_{(1 / 2) -, (1 / 2) +} =
    &\frac{2 (1 + \xi) }{3} \left[ H^{q T}_1 + H^{q T}_{16} + \frac{ (1 + \xi)}{4 (1 -
    \xi)} H^{q T}_3 \right]\\
    &- \frac{(1 - \xi)}{3} \left[ H^{q T}_2 + 2 H^{q T}_9 + H^{q T}_{10} - \left( \xi
    H^{q T}_{15} - \frac{1}{2} H^{q T}_{16} \right) + \frac{| C |^2}{8} \left(
    \frac{1}{2} H^{q T}_4 + H^{q T}_6 \right) \right]\\
    &- \frac{4 (1 - \xi^2)}{3} \left( \frac{1}{2} H^{q T}_9 + H^{q T}_{13} -
    \frac{1}{2} \xi H^{q T}_{14} \right) - \frac{2}{3} \left[ H^{q T}_2 - \left( \xi
    H^{q T}_{15} - \frac{1}{2} H^{q T}_{16} \right) \right]\\
    &+ \frac{7}{3} \xi H^{q T}_5 + \frac{1}{3}  (H^{q T}_5 + H^{q T}_7) - \frac{4 \xi}{3}
    \left( \xi H^{q T}_9 - \frac{1}{2} H^{q T}_{10} \right) + \frac{| C |^2}{3}
    \left[ H^{q T}_7 - \frac{5}{8} H^{q T}_8 - \frac{1}{2} \left( \xi H^{q T}_{11} +
    \frac{1}{2} H^{q T}_{12} \right) \right]\\
    &+ \frac{1}{3 (1 + \xi)} \left( \xi H^{q T}_{11} + \frac{1}{2} H^{q T}_{12} +
    \frac{| C |^2}{2} H^{q T}_8 \right) - \frac{4}{3 (1 - \xi)} (H^{q T}_5 + H^{q T}_7)
    + \frac{2 \xi}{3 (1 - \xi^2)} \left( H^{q T}_{11} - \frac{1}{2} H^{q T}_{12}
    \right)\\
    &- \frac{4 [D (1 - \xi^2) - \xi]}{3 (1 - \xi)} \left[ \frac{1}{2} H^{q T}_3 -
    H^{q T}_5 - \frac{1}{1 + \xi} \left( \frac{1}{2} \xi H^{q T}_{11} - \frac{3}{4}
    H^{q T}_{12} \right) \right]\\
    &+ \frac{2 [D (1 - \xi^2) + \xi]}{3 (1 - \xi^2)} \left[ \frac{\xi}{(1 -
    \xi)}  \left( \frac{1}{2} H^{q T}_4 - \xi H^{q T}_6 \right) + 2 H^{q T}_7 - (1 -
    \xi^2) H^{q T}_{16} + \frac{| C |^2}{4} H^{q T}_8 \right]\\
    &- \frac{2 [D (1 - \xi^2) - \xi] [D (1 - \xi^2) + \xi]}{3 (1 - \xi^2) (1 -
    \xi)} \left( \frac{1}{2} H^{q T}_4 - H^{q T}_6 - \frac{1}{1 + \xi} H^{q T}_8 \right),
  \end{split}
\end{equation}
\begin{equation}
  \begin{split}
    \mathcal{A}'^{\, q}_{(1 / 2) -, (- 1 / 2) +} =
    &\frac{2 \sqrt{1 - \xi^2}}{3} \left[ \frac{1}{2} H^{q T}_2 + H^{q T}_5 + H^{q T}_9 -
    \frac{1}{2} \xi H^{q T}_{10} + \frac{1}{2} H^{q T}_{12} - \frac{1}{2} \left( \xi
    H^{q T}_{15} + \frac{1}{2} H^{q T}_{16} \right) + \frac{| C |^2}{32} H^{q T}_8
    \right]\\
    &+ \frac{(1 + \xi^2)}{\sqrt{1 - \xi^2}} H^{q T}_7 + \frac{1}{\sqrt{1 - \xi^2}}
    \left[ \frac{\xi}{2} \left( \frac{1}{2} H^{q T}_4 - \xi H^{q T}_6 \right) - \left(
    \xi H^{q T}_{11} + \frac{1}{2} H^{q T}_{12} \right) \right]\\
    &- \frac{\xi (1 - \xi)}{3 \sqrt{1 - \xi^2}} (H^{q T}_6 - 2 H^{q T}_7 + H^{q T}_{12}) -
    \frac{\xi (1 - \xi)}{3 \sqrt{(1 - \xi^2)^3}} H^{q T}_8 + \frac{[D (1 - \xi^2)
    + \xi]}{3 \sqrt{1 - \xi^2}} (H^{q T}_6 - 2 H^{q T}_7 + H^{q T}_{12})\\
    &+ \frac{[D (1 - \xi^2) + \xi]}{3 \sqrt{(1 - \xi^2)^3}} H^{q T}_8 - \frac{[D (1
    - \xi^2) - \xi] [D (1 - \xi^2) + \xi]}{3 \sqrt{(1 - \xi^2)^3}} H^{q T}_8,
  \end{split}
\end{equation}
\begin{equation}
  \begin{split}
    \mathcal{A}'^{\, q}_{(1 / 2) -, (- 3 / 2) +} =
    \frac{(1 - \xi)}{2 \sqrt{3}} \left[ \frac{1}{4} \left( \frac{1}{2} H^{q T}_4 +
    H^{q T}_6 \right) - H^{q T}_7 - \frac{1}{2} \left( H^{q T}_{11} - \frac{1}{2} H^{q T}_{12}
    \right) \right] + \frac{1}{8 \sqrt{3}} H^{q T}_8 - \frac{[D (1 - \xi^2) -
    \xi]}{4 \sqrt{3}  (1 + \xi)} H^{q T}_8,
  \end{split}
\end{equation}

\begin{equation}
  \begin{split}
    \mathcal{A}'^{\, q}_{(- 1 / 2) -, (3 / 2) +} =
    &- \frac{2 (1 - \xi)}{\sqrt{3}} \left[ \left( H^{q T}_1 - \frac{1}{2} H^{q T}_2
    \right) + \frac{| C |^2}{4} H^{q T}_7 + \left( \xi H^{q T}_9 - \frac{1}{2}
    H^{q T}_{10} \right) \right]\\
    &- \frac{| C |^2 (1 + \xi)}{4 \sqrt{3}} \left[ \frac{1}{2} \left(
    \frac{1}{2} H^{q T}_4 - H^{q T}_6 \right) + \frac{1 - \xi}{1 + \xi} \left(
    H^{q T}_{11} - \frac{1}{2} H^{q T}_{12} \right) \right]
    + \frac{| C |^2}{8 \sqrt{3}} H^{q T}_8 + \frac{2}{\sqrt{3}} \left( \xi
    H^{q T}_{15} - \frac{1}{2} H^{q T}_{16} \right)\\
    &- \frac{4}{\sqrt{3} (1 + \xi)} \left[ \xi \left( \frac{1}{2} H^{q T}_3 - \xi
    H^{q T}_5 \right) - \frac{1}{2 (1 - \xi)} \left( \xi H^{q T}_{11} - \frac{1}{2}
    H^{q T}_{12} \right) \right]\\
    &- \frac{2 [D (1 - \xi^2) - \xi]}{ \sqrt{3} (1 - \xi^2)} \left[ \left( \xi
    H^{q T}_{11} - \frac{1}{2} H^{q T}_{12} \right) + \frac{\xi}{(1 + \xi)} \left(
    \frac{1}{2} H^{q T}_4 - \xi H^{q T}_6 \right) + (1 - \xi) \frac{| C |^2}{8}
    H^{q T}_8 \right],
  \end{split}
\end{equation}

\begin{equation}
  \begin{split}
    \mathcal{A}'^{\, q}_{(- 1 / 2) -, (1 / 2) +} =
    &- \frac{4 (1 - \xi)^2}{3 \sqrt{(1 - \xi^2)}} H^{q T}_1 - \frac{4 (1 -
    \xi)}{3 \sqrt{1 - \xi^2}} \left[ \frac{| C |^2}{4} \left( 3 H^{q T}_7 -
    H^{q T}_{12} + \frac{\xi (1 + \xi)}{1 - \xi} H^{q T}_7 \right) \right] +
    \frac{4}{3 \sqrt{1 - \xi^2}} H^{q T}_2\\
    &+ \frac{8 \xi}{3 \sqrt{1 - \xi^2}} \left[ - \left( \xi H^{q T}_{15} -
    \frac{1}{2} H^{q T}_{16} \right) - \frac{1 - \xi}{1 + \xi} \left( \frac{1}{2}
    H^{q T}_3 - \xi H^{q T}_5 \right) + \left( \xi H^{q T}_9 - \frac{1}{2} H^{q T}_{10}
    \right) \right]\\
    &+ \frac{2 | C |^2 \xi}{3 \sqrt{1 - \xi^2}} \left[ H^{q T}_3 - \frac{1}{2}
    \frac{1 + \xi}{1 - \xi} \left( \frac{1}{2} H^{q T}_4 - H^{q T}_6 \right) +
    H^{q T}_{11} \right]\\
    &+ \frac{8 \sqrt{1 - \xi^2}}{3} \left[ \xi \left( \xi H^{q T}_{13} -
    \frac{1}{2} H^{q T}_{14} \right) - \frac{| C |^2}{4} \frac{1 + \xi^2}{1 -
    \xi^2} H^{q T}_5 \right]\\
    &+ \frac{8 \sqrt{1 - \xi^2}}{3} \left[ - \frac{| C |^2}{8} \left( H^{q T}_2 +
    2 H^{q T}_5 + 2 \left( H^{q T}_9 - \frac{1}{2} \xi H^{q T}_{10} \right) - \left( \xi
    H^{q T}_{15} + \frac{1}{2} H^{q T}_{16} \right) \right) - \frac{| C |^4}{128}
    H^{q T}_8 \right]\\
    &- \frac{8 (1 - \xi)}{3 \sqrt{(1 - \xi^2)^3}} \left[ \frac{\xi}{1 - \xi}
    \left( \xi H^{q T}_{11} - \frac{1}{2} H^{q T}_{12} \right) + \frac{\xi (1 +
    \xi)}{(1 - \xi)}  \frac{| C |^2}{8} H^{q T}_8 \right]\\
    &+ \frac{8 [D (1 - \xi^2) - \xi]}{3 \sqrt{1 - \xi^2}} \left[ H^{q T}_1 - \left(
    \xi H^{q T}_{15} - \frac{1}{2} H^{q T}_{16} \right) - \frac{| C |^2}{8} \left(
    \frac{1 + \xi^2}{1 - \xi^2} H^{q T}_6 - 2 H^{q T}_7 + H^{q T}_{12} \right) \right]\\
    &+ \frac{8 [D (1 - \xi^2) - \xi]}{3 \sqrt{(1 - \xi^2)^3}} \left[ \xi (H^{q T}_3
    - 2 \xi H^{q T}_5) - \left( \xi H^{q T}_{11} - \frac{1}{2} H^{q T}_{12} \right) +
    \frac{| C |^2}{8} (\xi H^{q T}_4 - H^{q T}_8) \right]\\
    &+ \frac{8 [D (1 - \xi^2) + \xi] [D (1 - \xi^2) - \xi]}{3 \sqrt{(1 -
    \xi^2)^5}} \left[ \xi \left( \frac{1}{2} H^{q T}_4 - \xi H^{q T}_6 \right) +
    \frac{| C |^2 (1 - \xi^2)}{8} H^{q T}_8 \right],
  \end{split}
\end{equation}
\begin{equation}
  \begin{split}
    \mathcal{A}'^{\, q}_{(- 1 / 2) -, (- 1 / 2) +} =
    &- \frac{4 (1 - \xi^2)}{3} \left( H^{q T}_{13} - \frac{1}{2} \xi H^{q T}_{14}
    \right) + \frac{2 (1 - \xi)}{3} \left[ H^{q T}_1 - \frac{1 - \xi}{4 (1 + \xi)}
    H^{q T}_3 \right] - \frac{(1 - \xi)}{3 (1 + \xi)} H^{q T}_5\\
    &- \frac{(1 + \xi)}{3} \left[ H^{q T}_2 + 2 \xi H^{q T}_9 - H^{q T}_{10} - \left( \xi
    H^{q T}_{15} + \frac{1}{2} H^{q T}_{16} \right) - \frac{| C |^2}{8} \left(
    \frac{1}{2} H^{q T}_4 - H^{q T}_6 \right) \right]\\
    &- \frac{2}{3} \left[ H^{q T}_2 - \frac{1}{2} H^{q T}_7 + 2 H^{q T}_9 + \xi \left(
    \frac{7}{2} H^{q T}_5 - H^{q T}_{10} - H^{q T}_{15} - \frac{1}{2} H^{q T}_{16} \right)
    \right]\\
    &+ \frac{| C |^2}{3} \left[ H^{q T}_7 - \frac{5}{8} H^{q T}_8 - \frac{1}{2}
    \left( \xi H^{q T}_{11} + \frac{1}{2} H^{q T}_{12} \right) \right] - \frac{2}{3 (1
    + \xi)} \left( H^{q T}_5 + 2 H^{q T}_7 - \xi H^{q T}_{11} - \frac{| C |^2}{4} H^{q T}_8
    \right)\\
    &+ \frac{1}{3 (1 - \xi)} \left( \xi H^{q T}_{11} + \frac{1}{2} H^{q T}_{12} \right)
    + \frac{2 \xi}{3 (1 - \xi^2)} \left( \xi H^{q T}_{11} + \frac{1}{2} H^{q T}_{12}
    \right)\\
    &+ \frac{2 [D (1 - \xi^2) - \xi]}{3 (1 - \xi^2)} \left[ \frac{\xi}{(1 +
    \xi)} \left( \frac{1}{2} H^{q T}_4 - \xi H^{q T}_6 \right) + 2 H^{q T}_7 \right]\\
    &+ \frac{2 [D (1 - \xi^2) + \xi]}{3 (1 - \xi^2)} \left[ 2 (1 - \xi)  \left(
    \frac{1}{2} H^{q T}_3 + H^{q T}_5 \right) + \xi H^{q T}_{11} - \frac{3}{2} H^{q T}_{12} -
    (1 - \xi^2) H^{q T}_{16} + \frac{| C |^2}{4} H^{q T}_8 \right]\\
    &+ \frac{2 [D (1 - \xi^2) + \xi] [D (1 - \xi^2) - \xi]}{3 (1 - \xi^2) (1 +
    \xi)} \left( \frac{1}{2} H^{q T}_4 + H^{q T}_6 + \frac{H^{q T}_8}{1 - \xi} \right),
  \end{split}
\end{equation}

\begin{equation}
  \begin{split}
    \mathcal{A}'^{\, q}_{(- 1 / 2) -, (- 3 / 2) +} =
    &\frac{\sqrt{1 - \xi^2}}{\sqrt{3}} \left[ \frac{1 - \xi}{1 + \xi} \left(
    \frac{1}{2} H^{q T}_3 + H^{q T}_5 \right) + \frac{1}{2} H^{q T}_7 + \frac{1}{2}
    H^{q T}_{11} + \frac{| C |^2}{16} H^{q T}_8 \right] + \frac{(1 - \xi)}{\sqrt{3}
    \sqrt{1 - \xi^2}} \left( H^{q T}_7 - \frac{1}{2} H^{q T}_{12} \right)\\
    &+ \frac{\sqrt{1 - \xi^2} (1 - \xi)}{2 \sqrt{3}} \left( H^{q T}_{15} -
    \frac{1}{2} H^{q T}_{16} \right) + \frac{[D (1 - \xi^2) - \xi]}{2 \sqrt{3}
    \sqrt{1 - \xi^2}} \left[ \frac{1 - \xi}{1 + \xi} \left( \frac{1}{2} H^{q T}_4
    + H^{q T}_6 \right) + \frac{1}{1 + \xi} H^{q T}_8 \right],
  \end{split}
\end{equation}

\begin{equation}
  \begin{split}
    \mathcal{A}'^{\, q}_{(- 3 / 2) -, (3 / 2) +} =
    \frac{1}{\sqrt{1 - \xi^2}} \left[ \frac{\xi}{2} \left( \frac{1}{2} H^{q T}_4 -
    \xi H^{q T}_6 \right) + \left( \xi H^{q T}_{11} - \frac{1}{2} H^{q T}_{12} \right)
    \right] + \frac{| C |^2 \sqrt{1 - \xi^2}}{16} H^{q T}_8,
  \end{split}
\end{equation}
\begin{equation}
  \begin{split}
    \mathcal{A}'^{\, q}_{(- 3 / 2) -, (1 / 2) +} =
    &- \frac{2 (1 + \xi)}{\sqrt{3}} \left[ H^{q T}_1 - \frac{1}{2} H^{q T}_2 - \xi
    H^{q T}_9 + \frac{1}{2} {H^{q T}_{10}}^{} + \frac{| C |^2}{4} \left( H^{q T}_7 -
    \frac{1}{2} H^{q T}_{11} - \frac{1}{4} H^{q T}_{12} \right) \right]\\
    &+ \frac{| C |^2 (1 - \xi)}{8 \sqrt{3}} \left( \frac{1}{2} H^{q T}_4 + H^{q T}_6
    \right) + \frac{2}{\sqrt{3}} \left( \xi H^{q T}_{15} - \frac{1}{2} H^{q T}_{16} +
    \frac{| C |^2}{16} H^{q T}_8 \right)\\
    &- \frac{4 \xi}{\sqrt{3} (1 - \xi)} \left( \frac{1}{2} H^{q T}_3 - \xi H^{q T}_5
    \right) + \frac{2}{\sqrt{3} (1 - \xi^2)} \left( \xi H^{q T}_{11} - \frac{1}{2}
    H^{q T}_{12} \right)\\
    &- \frac{2 [D (1 - \xi^2) + \xi]}{\sqrt{3} (1 - \xi^2)} \left[ \frac{\xi}{1
    - \xi} \left( \frac{1}{2} H^{q T}_4 - \xi H^{q T}_6 \right) + \left( \xi H^{q T}_{11}
    - \frac{1}{2} H^{q T}_{12} \right) + \frac{| C |^2}{8 (1 + \xi)} H^{q T}_8
    \right],
  \end{split}
\end{equation}

\begin{equation}
  \begin{split}
    \mathcal{A}'^{\, q}_{(- 3 / 2) -, (- 1 / 2) +} =
    &+ \frac{4 (1 + \xi) \sqrt{1 - \xi^2}}{\sqrt{3}} \left[ \left( \xi H^{q T}_{13}
    - \frac{1}{2} H^{q T}_{14} \right) + \frac{| C |^2}{8} \left( H^{q T}_{15} +
    \frac{1}{2} H^{q T}_{16} \right) \right]\\
    &- \frac{2 \sqrt{1 - \xi^2}}{\sqrt{3}} \left( H^{q T}_1 - \frac{1}{1 - \xi}
    H^{q T}_2 + \frac{| C |^2}{2} \left( \frac{1}{2} H^{q T}_3 + H^{q T}_5 + \frac{1}{2}
    H^{q T}_7 \right) + \frac{| C |^4}{32} H^{q T}_8 \right)\\
    &- \frac{2 \xi}{\sqrt{3 (1 - \xi^2)}} \left[ (H^{q T}_3 - 2 \xi H^{q T}_5) +
    \frac{| C |^2}{4} \left( \frac{1}{2} H^{q T}_4 - \xi H^{q T}_6 \right) \right]\\
    &+ \frac{4 (1 + \xi)}{\sqrt{3 (1 - \xi^2)}} \left( \xi H^{q T}_9 - \frac{1}{2}
    H^{q T}_{10} - \frac{| C |^2}{4} H^{q T}_7 \right) + \frac{| C |^2}{2 \sqrt{3
    (1 - \xi^2)}} [H^{q T}_{11} + H^{q T}_{12}]\\
    &- \frac{4 [D (1 - \xi^2) + \xi]}{\sqrt{3 (1 - \xi^2)^3}} \left[ \left( \xi
    H^{q T}_{11} - \frac{1}{2} H^{q T}_{12} \right) \right]\\
    &- \frac{4 [D (1 - \xi^2) + \xi]}{\sqrt{3 (1 - \xi^2)}} \left[ \left( \xi
    H^{q T}_{15} - \frac{1}{2} H^{q T}_{16} \right) + \frac{| C |^2}{8} \left(
    \frac{1}{2} H^{q T}_4 + H^{q T}_6 + \frac{1}{1 - \xi} H^{q T}_8 \right) \right],
  \end{split}
\end{equation}
\begin{equation}
  \begin{split}
    \mathcal{A}'^{\, q}_{(- 3 / 2) -, (- 3 / 2) +} =
    &- (1 - \xi) \left( \frac{1}{2} H^{q T}_3 + H^{q T}_5 + \left( \xi H^{q T}_{15} -
    \frac{1}{2} H^{q T}_{16} \right) + \frac{| C |^2}{8} \left( \frac{1}{2}
    H^{q T}_4 + H^{q T}_6 \right) \right)\\
    &- \left( H^{q T}_7 + \frac{| C |^2}{8} H^{q T}_8 \right) - \frac{1}{(1 + \xi)}
    \left( \xi H^{q T}_{11} - \frac{1}{2} H^{q T}_{12} \right).
  \end{split}
\end{equation}
The other amplitudes with helicity flip $\mathcal{A}'^{\, q}_{i +, j -}$ can be obtained by the
parity invariance~\eqref{parityinvariancereal}.

In the forward limit, since $\bar{u}_{\alpha'}P^{\alpha'}=P^\alpha u_\alpha=0$,
the non-zero contributions of $H^{q T}_{1,2}(x,0,0)$ give the transversity distribution
function $h_1(x)$~\cite{Hoodbhoy:1998vm,Cosyn:2018rdm}. In addition, as discussed
above, there is no orbital contribution, i.e. $F(\zeta \neq 0)=0$, which constrains that the
only helicity conserved amplitudes,
$\mathcal{A}^q_{(1 / 2) -, (3 / 2) +}$, $\mathcal{A}^q_{(-3 / 2) -, (-1 / 2) +}$,
and $\mathcal{A}^q_{(-1 / 2) -, (1 / 2) +}$, remain finite.
Thus, the transversity distribution function $h_1 (x)$ can then be given as
\begin{equation}\label{quarktransversitydistribution}
    2 \, [H^{q T}_1(x,0,0)-H^{q T}_2(x,0,0)]=h_1(x),
\end{equation}
where the number density is from two components, spin-1/2 and spin-1, due to the fact that the
Rarita-Schwinger spinor is composed  by the fields of spin-1/2 and spin-1.
Compared the decompositions of spin-3/2 quark transversity GPDs \eqref{quarkdecompositions}
with those of spin-1/2 \cite{Diehl:2001pm} and of spin-1 \cite{Cosyn:2018rdm},
$H^{q T}_1(x,0,0)$ is from spin-1/2 component and $H^{q T}_2(x,0,0)$ from spin-1.
The minus sign comes from the definition difference between~\eqref{quarkde} in this work
and Eq.~(17) in \cite{Cosyn:2018rdm}.

Recall that the transversity distribution function $h_1(x)$ can be measured in semi-inclusive
deep inelastic scattering from transverse polarized targets in the scaling limit using the ``Collins asymmetry"~\cite{Collins:1992kk},
\begin{equation}
    A_\text{Coll}=\frac{\sum_q e_q^2 \, h_1^q \, \Delta_T^0 D_q^h}
    {\sum_q e_q^2 \, f_1^q \, D_q^h},
\end{equation}
where $e_q$, $f_1^q$, $\Delta_T^0 D_q^h$, and $D_q^h$ are the quark charge, the
unpolarized parton distribution function, the spin-dependent fragmentation function,
and the spin-independent fragmentation function, respectively.



\subsection{Gluon transversity GPDs}\label{subsectiongluon}

\quad\,\
According to the corresponding definition~\eqref{gluontransversitygpds}, the gluon
transversity GPDs at twist 2 can be obtained from
\begin{equation}
    T^{g i j}_{\lambda' \lambda} = -\bar{u}_{\alpha'} (p',\lambda')
    \mathcal{H}^{g T, i j, \alpha' \alpha}(x,\xi,t) u_\alpha(p,\lambda),
\end{equation}
where
\begin{align}\label{gluondecompositions}
        & \mathcal{H}^{g T,ij,\alpha' \alpha}= \notag \\
        &\hat{\bm{S}}_{ij}\left\{
        H_1^{g T} \frac{\left(\Delta^i + 2 \xi P^i \right) i \sigma^{nj}}
        {M \left( P \cdot n\right)} g^{\alpha' \alpha}
        + H_2^{g T} \frac{n^{[ \alpha' \nobracket} g^{\nobracket \alpha ] i}
        \left( \slashed{n}P^j - P\cdot n \gamma^j \right)}{\left( P \cdot n\right)^2}
        \right. \notag \\
        &\left.
        + H_3^{g T} \frac{\left(\Delta^i + 2 \xi P^i \right) n^{[ \alpha' \nobracket}
        g^{\nobracket \alpha ] j}}{M \left( P \cdot n\right)}
        +H_4^{g T} \frac{\left(\Delta^i + 2 \xi P^i \right)\left(\Delta^j
        + 2 \xi P^j \right) P^{\alpha'}P^\alpha}{M^5} \right. \notag \\
        &\left.
        + H_5^{g T} \frac{\left( P \cdot n g^{\alpha' i} -n^{\alpha'} P^i \right)
        \left( P \cdot n g^{\alpha j} -n^{\alpha} P^j \right)}{M \left( P \cdot n\right)^2}
        + H_6^{g T} \frac{\left( \Delta \cdot n g^{\alpha' i} -n^{\alpha'} \Delta^i \right)
        \left( \Delta \cdot n g^{\alpha j} -n^{\alpha}
        \Delta^j \right)}{M \left( P \cdot n\right)^2} \right. \notag \\
        &\left.
        + H_7^{g T} \frac{\left(\Delta^i + 2 \xi P^i \right)}{M}
        \frac{\left( \Delta \cdot n \, n^{\{ \alpha' \nobracket} g^{\nobracket \alpha \} j}
        -2 n^{\alpha'} n^\alpha \Delta^j \right)}{\left( P \cdot n\right)^2}
        + H_8^{g T} \frac{\left(\Delta^i + 2 \xi P^i \right)}{M}
        \frac{\left( P \cdot n \, n^{\{ \alpha' \nobracket} g^{\nobracket \alpha \} j}
        -2 n^{\alpha'} n^\alpha P^j \right)}{\left( P \cdot n\right)^2} \right. \notag \\
        &\left.
        + H_9^{g T}\frac{\left(\Delta^i + 2 \xi P^i \right)}{M^2}
        \frac{\left( \Delta \cdot n \, P^{[ \alpha' \nobracket} g^{\nobracket \alpha ] j}
        - P^{[ \alpha' \nobracket} n^{\nobracket \alpha ]}
        \Delta^j \right)}{M \left( P \cdot n\right)}
        + H_{10}^{g T}\frac{\left(\Delta^i + 2 \xi P^i \right)}{M^2}
        \frac{\left( P \cdot n \, P^{[ \alpha' \nobracket} g^{\nobracket \alpha ] j}
        - P^{[ \alpha' \nobracket} n^{\nobracket \alpha ]}
        P^j \right)}{M \left( P \cdot n\right)}\right. \notag \\
        &\left.
        + H_{11}^{g T} \frac{\left( P \cdot n g^{\alpha' i} -n^{\alpha'} P^i \right)\slashed{n}
        \left( P \cdot n g^{\alpha j} -n^{\alpha} P^j \right)}{\left( P \cdot n\right)^3}
        + H_{12}^{g T} \frac{\left( \Delta \cdot n g^{\alpha' i} -n^{\alpha'} \Delta^i \right)
        \slashed{n} \left( \Delta \cdot n g^{\alpha j} -n^{\alpha} \Delta^j \right)}
        {\left( P \cdot n\right)^3} \right. \notag \\
        & \left.
        + H^{g T}_{13} \left(\Delta^i + 2 \xi P^i \right)\slashed{n}
        \frac{\left( \Delta \cdot n n^{\{ \alpha' \nobracket} g^{\nobracket \alpha \} j}
        - 2 n^{\alpha'} n^\alpha \Delta^j \right)}{\left( P \cdot n\right)^3}
        + H^{g T}_{14} \left(\Delta^i + 2 \xi P^i \right)\slashed{n}
        \frac{\left( P \cdot n \, n^{\{ \alpha' \nobracket} g^{\nobracket \alpha \} j} - 2
        n^{\alpha'} n^\alpha P^j \right)}{\left( P \cdot n\right)^3} \right. \notag  \\
        & \left. + H^{g T}_{15} \left(\Delta^i + 2 \xi P^i \right)\slashed{n}
        \frac{\left( \Delta \cdot n \, P^{[ \alpha' \nobracket} g^{\nobracket \alpha ] j}
        - P^{[ \alpha' \nobracket} n^{\nobracket \alpha ]} \Delta^j \right) }
        {M^2 \left( P \cdot n\right)^2}+ H^{g T}_{16}
        \left(\Delta^i + 2 \xi P^i \right)\slashed{n}
        \frac{\left( P \cdot n \, P^{[ \alpha' \nobracket} g^{\nobracket \alpha ] j}
        - P^{[ \alpha' \nobracket} n^{\nobracket \alpha ]} P^j \right)}
        {M^2 \left( P \cdot n\right)^2} \right\},
\end{align}
where the operator $\hat{\bm{S}}_{i j}$ implies the symmetrization and removal of trace between indices $i$ and $j$,
and the variables $x$, $\xi$, $t$ in the gluon transversity GPDs $H^{g T}_i$ are omitted too.
Similarly, the relations in Eq.~\eqref{paritytime} constrains the even or odd behavior in terms of skewness $\xi$ as
\begin{equation}\label{timereversalgluon}
    \begin{split}
        H^{g T}_i(x,\xi,t) & =H^{g T}_i(x,-\xi,t) \quad \text{with}
        \quad i=1,3 \sim 7,10 \sim 13,16,\\
        H^{g T}_j(x,\xi,t) & =-H^{g T}_j(x,-\xi,t) \quad \text{with} \quad j=2,8,9,14,15,\\
    \end{split}
\end{equation}
where $H^{g T}_{2,8,9,14,15}$ are odd and others are even in $\xi$. Odd GPDs vanish at $\xi=0$.

We also attempt to give the tensor FFs of the local current
$G^{\mu \nu}(0) G^{\rho \sigma}(0)$ corresponding to the gluon transversity GPDs and
\begin{equation}
    T^{\mu \nu,\rho \sigma}= \frac{1}{M} \left\langle p', \lambda' \left|\text{Tr}
    \hat{\bm{S}}_{\nu \rho}[G^{\mu \nu}(0) G^{\rho \sigma}(0)] \right|p,\lambda \right\rangle
    =-2 \, \bar{u}_{\alpha'} (p',\lambda') \hat{\bm{S}}_{\nu \rho}
    \mathcal{F}_g^{\mu \nu, \rho \sigma, \alpha' \alpha} u_\alpha(p,\lambda).
\end{equation}
The gluon transversity GPDs in Eq.~\eqref{gluondecompositions} can determine the tensor
FFs $\mathcal{F}_g^{\mu \nu, \rho \sigma,\alpha' \alpha}$ by sum rules.
Eliminating the $n^\rho n^\sigma n^\delta$ and $n^\rho n^\sigma n^\delta n^\nu$ terms since
there are only two $n$'s in the operator $G^{n i}(0) G^{j n}(0)$ and considering the $\xi$ symmetry \eqref{timereversalgluon}, there remains 6 tensor FFs defined as
\begin{equation}
    \begin{split}
        \mathcal{F}_g^{\mu \nu, \rho \sigma,\alpha' \alpha,}=& \frac{P^{[ \mu \nobracket}
        \Delta^{\nobracket \nu ]}}{M^2} \left( G^{g T}_1(t)  i \sigma^{\sigma \rho}
        g^{\alpha' \alpha} + G^{g T}_3(t) g^{\sigma [ \alpha' \nobracket}
        g^{\nobracket \alpha ] \rho} + G^{g T}_4(t)
        \frac{P^{[ \sigma \nobracket} \Delta^{\nobracket \rho ]}}{M^2}
        + G^{g T}_{10}(t) \frac{P^{[\alpha' \nobracket} g^{\nobracket \alpha ]
        [\rho \nobracket} P^{\nobracket \sigma ]}}{M^2} \right) \\ & + G^{g T}_5(t)
        \frac{P^{[\mu \nobracket} g^{\nobracket \nu ] \alpha' }}{M}
        \frac{P^{[\sigma \nobracket} g^{\nobracket \rho ] \alpha}}{M}
        + G^{g T}_6(t) \frac{\Delta^{[\mu \nobracket} g^{\nobracket \nu ] \alpha' }}{M}
        \frac{\Delta^{[\sigma \nobracket} g^{\nobracket \rho ] \alpha}}{M},
    \end{split}
\end{equation}
and the sum rules collecting the tensor FFs and transversity GPDs are
\begin{equation}\label{gluonsumrules}
\begin{split}
        \int^1_{-1} d x \, H^{g T}_i(x,\xi,t)&=G^{g T}_i (t) \quad \text{with} \quad
        i=1,3 \sim 6,10,\\
    \int^1_{-1} d x \, H^{g T}_j(x,\xi,t)&=0 \quad \text{with}
    \quad j=2,7,8,9,11 \sim 16.
\end{split}
\end{equation}
The forward limit tells that the local current $G^{\mu \nu}(0) G^{\rho \sigma}(0)$
contains a global symmetry and contributes to the unknown gluon tensor charge
$G^{g T}_5 (0)$.

Analogous to the quark amplitudes with helicity flip, the ones of gluon can be derived
according to the definition in Eq.~\eqref{helicityamplitude}.
There is also a common factor $F(\zeta)$ defined in Eq.~\eqref{commonfactor} in the
expressions of the gluon amplitudes. Different from the quark case, the spin-1
gluon takes $\mu'\,(\mu)=\pm 1$ instead of $\pm \frac{1}{2}$.
Using the same conventions in Eq.~\eqref{helicityamplitudeconventions} and the definition
of the real amplitude~\eqref{helicityamplitudereal}, we have the 16 gluon amplitudes with helicity flip as
\begin{equation}
  \begin{split}
    \mathcal{A}'^{\, g}_{(3 / 2) -, (3 / 2) +} =&
    - \sqrt{\frac{1 + \xi}{1 - \xi}} \left( \xi H^{g T}_9 - \frac{1}{2} H^{g T}_{10}
    \right) - \sqrt{1 - \xi^2} (1 + \xi) \left( \xi H^{g T}_{15} - \frac{1}{2}
    H^{g T}_{16} \right) - \frac{| C |^2 \sqrt{1 - \xi^2}}{8} H^{g T}_4,
  \end{split}
\end{equation}

\begin{equation}
  \begin{split}
    \mathcal{A}'^{\, g}_{(3 / 2) -, (1 / 2) +} =& - \frac{(1 - \xi^2)}{2 \sqrt{3}}
    \left[ \left( H^{g T}_9 + \frac{1}{1 - \xi}
    H^{g T}_{10} \right) + (1 + \xi) \left( H^{g T}_{15} + \frac{1}{2} H^{g T}_{16}
    \right) - \frac{| C |^2}{8} H^{g T}_4 \right]\\
    & + \frac{[D (1 - \xi^2) + \xi]}{2 \sqrt{3} (1 - \xi)} H^{g T}_4,
  \end{split}
\end{equation}

\begin{equation}
  \begin{split}
    \mathcal{A}'^{\, g}_{(3 / 2) -, (- 1 / 2) +} =&
    \frac{\sqrt{1 - \xi^2}}{8 \sqrt{3}} H^{g T}_4 + \frac{\sqrt{1 - \xi^2} (1 +
    \xi)}{4 \sqrt{3}} \left( H^{g T}_9 + \frac{1}{2} H^{g T}_{10} \right) - \frac{[D
    (1 - \xi^2) + \xi]}{4 \sqrt{3}} \sqrt{\frac{1 + \xi}{1 - \xi}} H^{g T}_4,
  \end{split}
\end{equation}

\begin{equation}
  \begin{split}
    \mathcal{A}'^{\, g}_{(3 / 2) -, (- 3 / 2) +} = - \frac{1 - \xi^2}{16} H^{g T}_4,
  \end{split}
\end{equation}

\begin{equation}
  \begin{split}
    \mathcal{A}'^{\, g}_{(1 / 2) -, (3 / 2) +} =&
    - \frac{2 (1 - \xi^2)}{\sqrt{3}} \left[ H^{g T}_1 + \frac{1}{2} H^{g T}_2 - \left(
    \frac{1}{2} H^{g T}_{11} + 2 \xi H^{g T}_{12} \right) + \frac{| C |^4}{32} H^{g T}_4
    \right]\\ &
    - \frac{4 (1 - \xi^2) (1 - \xi)}{\sqrt{3}} \left[ \left( \xi H^{g T}_{13} -
    \frac{1}{2} H^{g T}_{14} \right) + \frac{| C |^2}{8} \left( H^{g T}_{15} -
    \frac{1}{2} H^{g T}_{16} \right) \right]\\ &
    + \frac{4 (1 - \xi)}{\sqrt{3}} \left( \frac{1}{2} H^{g T}_3 - \xi H^{g T}_7 +
    \frac{1}{2} H^{g T}_8 \right) + \frac{4}{\sqrt{3}} \left( \frac{1}{4} H^{g T}_5 +
    \xi H^{g T}_6 \right) - \frac{| C |^2}{2 \sqrt{3}} [(1 + \xi^2) H^{g T}_9 -
    H^{g T}_{10}]\\ &
    - \frac{4 [D (1 - \xi^2) - \xi]}{\sqrt{3} } \left[ \left( \xi H^{g T}_{15} -
    \frac{1}{2} H^{g T}_{16} \right) + \frac{1}{(1 - \xi^2)} \left( \xi H^{g T}_9 -
    \frac{1}{2} H^{g T}_{10} \right) + \frac{| C |^2}{8 (1 + \xi)} H^{g T}_4 \right],
  \end{split}
\end{equation}

\begin{equation}
  \begin{split}
    \mathcal{A}'^{\, g}_{(1 / 2) -, (1 / 2) +} = &
    \frac{2 \sqrt{1 - \xi^2} (1 + \xi)}{3} \left( H^{g T}_1 + \frac{1}{2} H^{g T}_2
    \right) - \frac{\sqrt{1 - \xi^2} (1 - \xi)}{3} \left( H^{g T}_3 - 2 H^{g T}_6 + 2
    H^{g T}_7 + H^{g T}_8 - \xi H^{g T}_{15} - \frac{1}{2} H^{g T}_{16} \right)\\ &
    - \frac{2 \sqrt{1 - \xi^2}}{3} \left[ H^{g T}_3 + \frac{1}{2} H^{g T}_5 + 2 H^{g T}_7
    - \xi H^{g T}_8 - \xi \left( H^{g T}_{15} - \frac{1}{2} H^{g T}_{16} \right) + \frac{5
    | C |^2}{16} \left( H^{g T}_4 + \frac{4}{5} \xi H^{g T}_9 + \frac{2}{5} H^{g T}_{10}
    \right) \right]\\ &
    + \frac{2 \sqrt{(1 - \xi^2)^3}}{3} \left( H^{g T}_7 - \frac{1}{4} H^{g T}_{11} +
    H^{g T}_{12} - 2 H^{g T}_{13} + \xi H^{g T}_{14} \right)\\ &
    + \frac{(1 - \xi)}{3 \sqrt{1 - \xi^2}} \left[ \left( \xi H^{g T}_9 +
    \frac{1}{2} H^{g T}_{10} \right) + \frac{2 \xi}{(1 - \xi)} \left( H^{g T}_9 -
    \frac{1}{2} H^{g T}_{10} \right) + \frac{1 + \xi}{2 (1 - \xi)} | C |^2 H^{g T}_4
    \right]\\ &
    + \frac{2 [D (1 - \xi^2) - \xi]}{3 \sqrt{(1 - \xi^2)}} \left( \xi H^{g T}_9 -
    \frac{3}{2} H^{g T}_{10} + \frac{| C |^2}{4} H^{g T}_4 \right) - \frac{2
    \sqrt{(1 - \xi^2)} [D (1 - \xi^2) - \xi]}{3} H^{g T}_{16}\\ &
    + \frac{2 [D (1 - \xi^2) + \xi] [D (1 - \xi^2) - \xi]}{3 \sqrt{(1 -
    \xi^2)^3}} H^{g T}_4,
  \end{split}
\end{equation}

\begin{equation}
  \begin{split}
    \mathcal{A}'^{\, g}_{(1 / 2) -, (- 1 / 2) +} = &
    \frac{1 - \xi^2}{3}  \left[ H^{g T}_3 + \frac{1}{4} H^{g T}_5 - H^{g T}_6 + 2 H^{g T}_7 -
    \xi H^{g T}_8 - \left( \xi H^{g T}_{15} + \frac{1}{2} H^{g T}_{16} \right) + \frac{|
    C |^2}{16} H^{g T}_4 \right]\\ &
    - \left[ \frac{1}{3} H^{g T}_4 + \left( \xi H^{g T}_9 + \frac{1}{2} H^{g T}_{10}
    \right) \right] + \frac{1}{3 (1 - \xi)} (H^{g T}_4 + (1 - \xi^2) H^{g T}_{10})\\ &
    + \frac{[D (1 - \xi^2) - \xi]}{3 (1 - \xi^2)} (H^{g T}_4 + (1 - \xi^2)
    H^{g T}_{10}) - \frac{[D (1 - \xi^2) + \xi] [D (1 - \xi^2) - \xi]}{3 (1 -
    \xi^2)} H^{g T}_4,
  \end{split}
\end{equation}

\begin{equation}
  \begin{split}
    \mathcal{A}'^{\, g}_{(1 / 2) -, (- 3 / 2) +} = \frac{\sqrt{1 - \xi^2}}{8
    \sqrt{3}}  [H^{g T}_4 - (1 - \xi) (2 H^{g T}_9 - H^{g T}_{10})] - \frac{[D (1 - \xi^2)
    - \xi]}{4 \sqrt{3}} \sqrt{\frac{1 - \xi}{1 + \xi}} H^{g T}_4,
  \end{split}
\end{equation}

\begin{equation}
  \begin{split}
    \mathcal{A}'^{\, g}_{(- 1 / 2) -, (3 / 2) +} = &
    - \frac{2 | C |^2  (1 - \xi) \sqrt{1 - \xi^2}}{\sqrt{3}} \left[ \left(
    H^{g T}_1 - \frac{1}{2} H^{g T}_3 + \xi H^{g T}_7 - \frac{1}{2} H^{g T}_8 \right) +
    \frac{| C |^2}{8}  \left( H^{g T}_9 - \frac{1}{2} H^{g T}_{10} \right) \right]\\ &
    + \frac{2 \sqrt{1 - \xi^2}}{\sqrt{3}} \left[ \left( \frac{2 \xi}{1 + \xi}
    H^{g T}_2 - H^{g T}_{11} - 4 \xi^2 H^{g T}_{12} \right) + \frac{| C |^2}{4} (H^{g T}_5 +
    4 \xi H^{g T}_6 + 4 \xi H^{g T}_{15} - 2 H^{g T}_{16}) + \frac{| C |^4}{16} (H^{g T}_4)
    \right]\\ &
    - \frac{2}{\sqrt{3 (1 - \xi^2)}} \left( H^{g T}_5 + 4 \xi^2 H^{g T}_6 - | C |^2
    \left( \xi H^{g T}_9 - \frac{1}{2} H^{g T}_{10} \right) \right)\\ &
    - \frac{2 | C |^2  [D (1 - \xi^2) - \xi]}{\sqrt{3 (1 - \xi^2)}} \left(
    \xi H^{g T}_9 - \frac{1}{2} H^{g T}_{10} \right) - \frac{| C |^4 [D (1 - \xi^2)
    - \xi]}{4 \sqrt{3}} \sqrt{\frac{1 - \xi}{1 + \xi}} H^{g T}_4,
  \end{split}
\end{equation}

\begin{equation}
  \begin{split}
    \mathcal{A}'^{\, g}_{(- 1 / 2) -, (1 / 2) +} = &
    \frac{4 (1 - \xi^2)}{3} \left( H^{g T}_1 - H^{g T}_6 - 2 H^{g T}_7 + \frac{1}{2}
    H^{g T}_{11} + 2 \xi^2 H^{g T}_{13} - \xi H^{g T}_{14} \right)\\ &
    - \frac{| C |^2 (1 - \xi^2)}{3} \left[  \left( H^{g T}_3 + \frac{1}{4} H^{g T}_5
    - H^{g T}_6 + 2 H^{g T}_7 - \xi H^{g T}_8 - \xi H^{g T}_{15} - \frac{1}{2} H^{g T}_{16}
    \right) + \frac{| C |^2}{16} H^{g T}_4 \right]\\ &
    - \frac{8}{3} \xi H^{g T}_2 + \frac{4}{3} H^{g T}_3 + H^{g T}_5 + \frac{4}{3} H^{g T}_6 +
    \frac{8}{3} H^{g T}_7 - \frac{4}{3} \xi H^{g T}_8 - \frac{8}{3} \left( \xi
    H^{g T}_{15} - \frac{1}{2} H^{g T}_{16} \right)\\ &
    + \frac{| C |^2}{3} \left( \frac{\xi}{1 + \xi} H^{g T}_4 + 2 \xi H^{g T}_9
    \right)\\ &
    - \frac{8 (1 + \xi)}{3} \left[ H^{g T}_1 - \frac{| C |^2}{8} H^{g T}_{10} - \xi
    H^{g T}_{15} + \frac{1}{2} H^{g T}_{16} \right] + \frac{4 \xi}{3 (1 - \xi^2)} [2
    \xi H^{g T}_9 - H^{g T}_{10}]\\ &
    - \frac{4 [D (1 - \xi^2) + \xi]}{3 (1 - \xi^2)} \left( 2 \xi H^{g T}_9 -
    H^{g T}_{10} + \frac{| C |^2}{4} H^{g T}_4 \right)\\ &
    + \frac{8 [D (1 - \xi^2) + \xi]}{3} \left( H^{g T}_1 - \frac{| C |^2}{8}
    H^{g T}_{10} - \xi H^{g T}_{15} + \frac{1}{2} H^{g T}_{16} \right)\\ &
    + \frac{| C |^2 [D (\xi^2 - 1) + \xi] [D (\xi^2 - 1) - \xi]}{3 (1 -
    \xi^2)} H^{g T}_4,
  \end{split}
\end{equation}

\begin{equation}
  \begin{split}
    \mathcal{A}'^{\, g}_{(- 1 / 2) -, (- 1 / 2) +} = &
    \frac{2 (1 - \xi^2)}{3} \left[ \sqrt{\frac{1 - \xi}{1 + \xi}} \left( H^{g T}_1
    - \frac{1}{2} H^{g T}_2 \right) + \sqrt{\frac{1 + \xi}{1 - \xi}} H^{g T}_6 \right]
    + \frac{2 \sqrt{(1 - \xi^2)^3}}{3} \left( H^{g T}_7 - \frac{1}{4} H^{g T}_{11} +
    H^{g T}_{12} - 2 H^{g T}_{13} + \xi H^{g T}_{14} \right)\\ &
    - \frac{\sqrt{1 - \xi^2} (1 + \xi)}{3} \left( H^{g T}_3 + 2 H^{g T}_7 - H^{g T}_8 -
    \xi H^{g T}_{15} - \frac{1}{2} H^{g T}_{16} \right)\\ &
    - \frac{2 \sqrt{(1 - \xi^2)}}{3} \left[ H^{g T}_3 + \frac{1}{2} H^{g T}_5 - \xi
    \left( H^{g T}_{15} + \frac{1}{2} H^{g T}_{16} \right) + (2 H^{g T}_7 - \xi H^{g T}_8 )
    \right]\\ &
    - \frac{| C |^2 \sqrt{(1 - \xi^2)}}{6} \left[ \left( \xi H^{g T}_9 +
    \frac{1}{2} H^{g T}_{10} + \frac{5}{4} H^{g T}_4 \right) \right]\\ &
    + \frac{(1 + \xi)}{3 \sqrt{1 - \xi^2}} \left[ \xi H^{g T}_9 + \frac{1}{2}
    H^{g T}_{10} + \frac{| C |^2}{2} H^{g }_4 + \frac{2 \xi}{1 + \xi} \left(
    H^{g T}_9 - \frac{5}{2} H^{g T}_{10} \right) \right]\\ &
    - \frac{[D (1 - \xi^2) - \xi]}{\sqrt{1 - \xi^2}} \left( H^{g T}_{10} - \frac{|
    C |^2}{6} H^{g }_4 \right) + \frac{2 [D (1 - \xi^2) + \xi]}{3 \sqrt{1 -
    \xi^2}} (\xi H^{g T}_9 - (1 - \xi^2) H^{g T}_{16})\\ &
    + \frac{2 [D (1 - \xi^2) + \xi] [D (1 - \xi^2) - \xi]}{3 \sqrt{(1 -
    \xi^2)^3}} H^{g T}_4,
  \end{split}
\end{equation}

\begin{equation}
  \begin{split}
    \mathcal{A}'^{\, g}_{(- 1 / 2) -, (- 3 / 2) +} = &
    \frac{(1 - \xi^2) (1 - \xi)}{2 \sqrt{3}} \left[ \left( H^{g T}_{15} -
    \frac{1}{2} H^{g T}_{16} \right) + \frac{1}{1 - \xi} \left( H^{g T}_9 + \frac{|
    C |^2}{8} H^{g T}_4 \right) - \frac{1}{1 - \xi^2} H^{g T}_{10} \right] +
    \frac{[D (1 - \xi^2) - \xi]}{2 \sqrt{3} (1 + \xi)} H^{g T}_4,
  \end{split}
\end{equation}

\begin{equation}
  \begin{split}
    \mathcal{A}'^{\, g}_{(- 3 / 2) -, (3 / 2) +} = - (H^{g T}_5 + 4 \xi^2
    H^{g T}_6) + | C |^2 \left( \xi H^{g T}_9 - \frac{1}{2} H^{g T}_{10} \right) +
    \frac{| C |^4 (1 - \xi^2)}{16} H^{g T}_4,
  \end{split}
\end{equation}

\begin{equation}
  \begin{split}
    \mathcal{A}'^{\, g}_{(- 3 / 2) -, (1 / 2) +} = &
    - \frac{2 | C |^2 \sqrt{1 - \xi^2} (1 + \xi)}{\sqrt{3}} \left[ H^{g T}_1 -
    \frac{1}{2} H^{g T}_3 - \xi H^{g T}_7 + \frac{1}{2} H^{g T}_8 - \frac{| C |^2}{8}
    \left( H^{g T}_9 + \frac{1}{2} H^{g T}_{10} \right) \right]\\ &
    - \frac{2 \sqrt{1 - \xi^2}}{\sqrt{3}} \left[ - \frac{2 \xi}{1 - \xi} H^{g T}_2
    + H^{g T}_{11} + 4 \xi^2 H^{g T}_{12} - | C |^2  \left( \frac{1}{4} H^{g T}_5 - \xi
    H^{g T}_6 + \xi H^{g T}_{15} - \frac{1}{2} H^{g T}_{16} \right) - \frac{| C |^4}{16}
    H^{g T}_4 \right]\\ &
    - \frac{2}{\sqrt{3 (1 - \xi^2)}} \left[ H^{g T}_5 + 4 \xi^2 H^{g T}_6 - | C |^2
    \left( \xi H^{g T}_9 - \frac{1}{2} H^{g T}_{10} \right) \right]\\ &
    - \frac{2 | C |^2 [D (1 - \xi^2) + \xi]}{\sqrt{3 (1 - \xi^2)}} \left[
    \left( \xi H^{g T}_9 - \frac{1}{2} H^{g T}_{10} \right) + (1 + \xi)  \frac{| C
    |^2}{8} H^{g T}_4 \right],
  \end{split}
\end{equation}

\begin{equation}
  \begin{split}
    \mathcal{A}'^{\, g}_{(- 3 / 2) -, (- 1 / 2) +} = &
    \frac{4 (1 - \xi^2) (1 + \xi)}{\sqrt{3}} \left( \xi H^{g T}_{13} - \frac{1}{2}
    H^{g T}_{14} + \frac{| C |^2}{8} \left( H^{g T}_{15} + \frac{1}{2} H^{g T}_{16}
    \right) \right)\\ &
    - \frac{2 (1 - \xi^2)}{\sqrt{3}} \left( \left( H^{g T}_1 - \frac{1}{2} H^{g T}_2
    \right) - \frac{1}{2} H^{g T}_{11} + 2 \xi H^{g T}_{12} + \frac{| C |^2}{4}
    H^{g T}_9 + \frac{| C |^4}{32} H^{g T}_4 \right)\\ &
    + \frac{2 (1 + \xi)}{\sqrt{3}} (H^{g T}_3 + 2 \xi H^{g T}_7 - H^{g T}_8) +
    \frac{1}{\sqrt{3}} \left[ (H^{g T}_5 - 4 \xi H^{g T}_6) + | C |^2 \left( H^{g T}_9 +
    \frac{1}{2} H^{g T}_{10} \right) \right]\\ &
    - \frac{4 [D (1 - \xi^2) + \xi]}{\sqrt{3}} \left[ \frac{1}{1 - \xi^2}
    \left( \xi H^{g T}_9 - \frac{1}{2} H^{g T}_{10} \right) + \left( \xi H^{g T}_{15} -
    \frac{1}{2} H^{g T}_{16} \right) + \frac{| C |^2}{8 (1 - \xi)} H^{g T}_4 \right],
  \end{split}
\end{equation}

\begin{equation}
  \begin{split}
    \mathcal{A}'^{\, g}_{(- 3 / 2) -, (- 3 / 2) +} =&
    - \frac{1 - \xi}{\sqrt{1 - \xi^2}} \left( \xi H^{g T}_9 - \frac{1}{2} H^{g T}_{10}
    \right) - \sqrt{1 - \xi^2} (1 - \xi) \left( \xi H^{g T}_{15} - \frac{1}{2}
    H^{g T}_{16} \right) - \frac{| C |^2  \sqrt{1 - \xi^2}}{8} H^{g T}_4.
  \end{split}
\end{equation}
The other amplitudes $\mathcal{A}'^{\, g}_{i +,j -}$ can be obtained from
the relation~\eqref{parityinvariancereal}.

In the forward limit, similar to the discussions on quark in
Sec.~\ref{subsectionquark}, there is no orbital contribution, i.e. $F(\zeta \neq 0)=0$, and
the only helicity conserved amplitudes $\mathcal{A}^g_{(- 1 / 2) -, ( 3 / 2) +}$ and
$\mathcal{A}^g_{(-3 / 2) -, ( 1 / 2) +}$ survive. The transversity distribution
function of the gluon is then obtained as
\begin{equation}\label{gluontransversitydistribution}
    -2 \left[H^{g T}_{5} (x,0,0)+ H^{g T}_{11} (x,0,0)\right] = x \Delta (x),
\end{equation}
defined in Refs.~\cite{Jaffe:1989xy,Hoodbhoy:1998vm} and $\Delta(x)$ has the similar
interpretation with $h_1(x)$ in Eq.~\eqref{quarktransversitydistribution}.
Since $(P \cdot n) \bar{u} (p,\lambda') u (p, \lambda) = M \bar{u} (p,\lambda')
\slashed{n} u (p, \lambda)$ in Eq.~\eqref{gluondecompositions} is satisfied in the forward limit, $H^{g T}_{5} (x,0,0)$
and $H^{g T}_{11} (x,0,0)$ correspond to the same tensor structure and all the gluon
transversity distributions are given by the spin-1 component. There is an obvious reason:
the helicity flip of gluon leads to a change of two units of angular momentum, which
is impossible for the spin-1/2 component. The minus sign in
Eq.~\eqref{gluontransversitydistribution} is explained by the same reason as in
Eq.~\eqref{quarktransversitydistribution}.

\section{Summary and discussion}\label{sectionsummary}

\quad\,
In this work, the quark and gluon transversity GPDs of spin-3/2 particles are respectively
derived and given for the first time. We find that there are 16 independent transversity
GPDs for each parton, constrained by the parity invariance. The even or odd property with respect to $\xi$
of the obtained transversity GPDs can help us to determine the corresponding tensor FFs that contain two tensor charges, where the quark tensor charge may describe
the electric dipole moment from 
the quark beyond the Standard Model and the gluon tensor charge is unknown. We need more study for their
physical meanings. There are seven tensor FFs for the local quark current and six for gluon
from the sum rules connecting the transversity GPDs and the tensor FFs.

The amplitudes with helicity flip in terms of transversity GPDs are obtained.
We conclude that all the amplitudes, including helicity non-flip and flip,
have the common factor $F(\zeta)$, which carries the whole complex part and represents the
transfer of the orbital angular momentum. In the forward limit, no orbital contribution, i.e.
$F(\zeta \neq 0)=0$, constrains the only helicity conserved amplitudes, which give the
corresponding parton distributions, to exist and the amplitudes with helicity flip
give the transversity distributions of each parton. Following the Rarita-Schwinger
spinor and our explanations, the transversity distribution of the parton in the spin-3/2
system is comprised of the spin-1/2 and spin-1 system, nevertheless the gluon transversity
distribution only includes the spin-1 since the gluon helicity flip changes two units
of angular momentum and
it leads to the violation of the angular momentum conversation in the
spin-1/2 system.

Finally, the present study of the transversity GPDs will be applied to a specific spin-3/2
particle, such as $\Delta$ resonance and $\Omega$ hyperon.
Such GPDs and also $N \to \Delta$ transition GPDs
could be investigated experimentally.
The timelike GPDs of $\Delta$ and $\Omega$ could be measured
by the two-photon processes at BESIII and KEK-B,
and the $N \to \Delta$ transition GPDs could be
measured at JLab and J-PARC.
In addition, the $\Delta$ could exist in nuclei as a small component,
so that its GPDs could become important for studying specific polarization observables.


\section*{Acknowledgments}
\quad\,\
We are grateful to Wim Cosyn for the very useful answer about the gluon transversity GPDs
for the spin-1 particle. 
We are also very grateful for the support of the Southern Center for Nuclear-Science Theory (SCNT), Institute
of Modern Physics.
This work is supported by the National Key Research and Development
Program of China under Contracts No. 2020YFA0406300, and by the National Natural Science
Foundation of China under Grants Nos. 11975245 and 12375142. This work is also supported by
the Sino-German CRC 110 Symmetries and the Emergence of Structure in QCD project by NSFC under Grant No. 12070131001, and the Key Research Program of Frontier Sciences,
CAS, under Grant No. Y7292610K1.
This work was partially supported by
the CAS president's international fellowship initiative
(Grant No. 2022VMA0003)
and by the Japan Society for the Promotion
of Science (JSPS) Grants-in-Aid for Scientific Research (KAKENHI)
Grant Number 19K03830.

\appendix

\renewcommand\thesection{Appendix~\Alph{section}}
\section{The Rarita-Schwinger Spinor}\label{appendixconventions}
\vspace{0.2cm} \par\noindent\par\setcounter{equation}{0}
\renewcommand{\theequation}{A\arabic{equation}}

The explicit form of the Rarita-Schwinger spinor of a spin-3/2 particle employed in
our work is \cite{Rarita:1941mf,Lurie:1968zz,Penrose:1985bww,Fu:2022bpf}
\begin{equation}\label{RSspinor}
u^{\alpha} (p, \lambda) = \sum_{\rho,\sigma} C^{\frac{3}{2} \lambda}_{1 \rho,
\frac{1}{2}\sigma} \epsilon^{\alpha} (p, \rho) u (p, \sigma),
\end{equation}
where the coefficient in Eq.~\eqref{RSspinor} is the
Clebsch-Gordan coefficient. The explicit light-front form expressions of the spin-1~\cite{Keister:1991sb} and -1/2~\cite{Lorce:2017isp} respectively are
\begin{subequations}\label{vectora2}
    \begin{align}
      \epsilon^{\alpha} (p,0) & = \frac{1}{M} \left(
      p^+,p^- - \frac{2 M^2}{p^+},\boldsymbol{\epsilon}_\perp (p,0) \right)^\text{T} \quad \text{with} \quad \boldsymbol{\epsilon}_\perp (p,0)=(p_1, p_2), \\
      \epsilon^{\alpha} (p,+1) & = - \left(
      0, \frac{\sqrt{2} (p_1 + i p_2)}{p^+} , \boldsymbol{\epsilon}_\perp (p,+1) \right)^\text{T} \quad \text{with} \quad \boldsymbol{\epsilon}_\perp (p,+1)=(\frac{1}{\sqrt{2}}, \frac{i}{\sqrt{2}}),\\
      \epsilon^{\alpha} (p,-1) & = \left(0, \frac{\sqrt{2} (p_1 - i p_2)}{p^+}, \boldsymbol{\epsilon}_\perp (p,-1) \right)^\text{T} \quad \text{with} \quad \boldsymbol{\epsilon}_\perp (p,-1)= (\frac{1}{\sqrt{2}}, \frac{- i}{\sqrt{2}}),
    \end{align}
\end{subequations}
and
\begin{equation}\label{diraca3}
    u (p, \sigma) = \frac{\left( \slashed{p} +
    M \right)}{\sqrt{2 p \cdot n}}
 \slashed{n} \chi_{\sigma},
\end{equation}
where $\chi_{\sigma}$ is the rest frame spinor.
The Rarita-Schwinger spinor in Eq.~\eqref{RSspinor} satisfies the Rarita-Schwinger
equation, as well as the subsidiary constraint equations,
\begin{equation}
    \left(\slashed{p}-M\right) u^{\alpha} (p, \lambda)=0,\quad
    \gamma_\alpha  u^{\alpha} (p, \lambda)=0, \quad
    p_\alpha  u^{\alpha} (p, \lambda)=0.
\end{equation}

\renewcommand\thesection{Appendix~\Alph{section}}
\section{On-Shell Identities}\label{appendixidentities}
\vspace{0.2cm} \par\noindent\par\setcounter{equation}{0}
\renewcommand{\theequation}{B\arabic{equation}}

Many identities and on-shell identities are proved and listed in
Refs.~\cite{Cotogno:2019vjb,Fu:2022rkn,Fu:2022bpf} and they will not be shown
in the following. Some particular and necessary identities are proved here.
Here we use the abbreviation, $\mathscr{P}^\mu=2 P^\mu-2 M \gamma^\mu$, to obtain the
more concise form.

From Eqs. (B13) and (B23C) in Ref.~\cite{Cotogno:2019vjb},
\begin{equation}
    g^{\alpha' \alpha} i \sigma^{\mu \nu} + g^{\mu \alpha'} g^{\alpha \nu} - g^{\mu \alpha} g^{\alpha' \nu} \doteq i \epsilon^{\alpha' \mu \nu \alpha} \gamma^5,
\end{equation}
and $\gamma^5 \doteq \frac{\slashed{\Delta} \gamma^5}{2 M}$ where $\doteq$ indicates on-shell equality,
one can derive
\begin{equation}
    g^{\alpha' \alpha} i \sigma^{\mu \nu} + g^{\mu [ \alpha' \nobracket} g^{\nobracket \alpha ] \nu} = g^{\alpha' \alpha} i \sigma^{\mu \nu} + g^{\mu \alpha'} g^{\alpha \nu} - g^{\mu \alpha} g^{\alpha' \nu} \doteq \frac{i \epsilon^{\alpha' \mu \nu \alpha}\slashed{\Delta} \gamma^5}{2 M}.
\end{equation}
In the forward limit, $\Delta=0$, and then
\begin{equation}
    g^{\alpha' \alpha} i \sigma^{\mu \nu} + g^{\mu [ \alpha' \nobracket} g^{\nobracket \alpha ] \nu} \doteq 0.
\end{equation}

There is an equality, instead of on-shell identity, from Schouten identity~\cite{Meissner:2009ww},
\begin{equation}\label{determination0}
    \det \begin{vmatrix}
     g^{\alpha \alpha'} & g^{\beta \alpha'} & g^{\gamma \alpha'} & g^{\delta
     \alpha'} & g^{\varepsilon \alpha'}\\
     g^{\alpha \nu} & g^{\beta \nu} & g^{\gamma \nu} & g^{\delta \nu} &
     g^{\varepsilon \nu}\\
     g^{\alpha \rho} & g^{\beta \rho} & g^{\gamma \rho} & g^{\delta \rho} &
     g^{\varepsilon \rho}\\
     g^{\alpha \sigma} & g^{\beta \sigma} & g^{\gamma \sigma} & g^{\delta
     \sigma} & g^{\varepsilon \sigma}\\
     g^{\alpha \tau} & g^{\beta \tau} & g^{\gamma \tau} & g^{\delta \tau} &
     g^{\varepsilon \tau}
   \end{vmatrix} = 0,
\end{equation}
where if the indices $\alpha'$ and $\alpha$ are coincide with ones of the initial and final
states depend on if the equality is contracted with states. Contracting this determinant with
\begin{equation}\label{contact}
     P_{\beta}, \, \Delta_{\gamma}, \, n_{\delta}, \, g_{\varepsilon}^i, \,
     P_{\nu}, \, \Delta_{\rho}, \, n_{\sigma} \, \text{and} \, g_{\tau}^j,
\end{equation}
one can obtain
\begin{equation}\label{determinationnogamma}
    \det \begin{vmatrix}
     g^{\alpha \alpha'} & P^{\alpha'} & \Delta^{\alpha'} & n^{\alpha'} & g^{i
     \alpha'}\\
     P^{\alpha} & P^2 & P \cdot \Delta & P \cdot n & P^i\\
     \Delta^{\alpha} & P \cdot \Delta & \Delta^2 & \Delta \cdot n &
     \Delta^i\\
     n^{\alpha} & P \cdot n & \Delta \cdot n & n \cdot n & n^i\\
     g^{\alpha j} & P^j & \Delta^j & n^j & g^{i j}
   \end{vmatrix} = 0.
\end{equation}
Using the on-shell identities which have been shown in
Refs.~\cite{Cotogno:2019vjb,Fu:2022rkn,Fu:2022bpf} and removing the trace term like $g^{i j}$, we replace the $\left( \Delta^i + 2 \xi P^i \right)\left( \Delta^j + 2 \xi P^j \right) g^{\alpha' \alpha}$
term by other terms surviving in Eq.~\eqref{gluondecompositions}. Similarly,
the product between $\slashed{n}$ and Eq.~\eqref{determinationnogamma} explains that $\left( \Delta^i + 2 \xi P^i \right)\left( \Delta^j + 2 \xi P^j \right) \slashed{n} g^{\alpha' \alpha}$ can also be eliminated.

In addition to the vector, the matrix like $\gamma^\mu$ and $\sigma^{\mu \nu}$ is permitted to
contract with Eq.~\eqref{determination0}. However, it is convenient and safe to contract
only one matrix with one of the lows or ranks because of the non-commutation property.
One can respectively replace $g^i_\varepsilon$, $g^i_\varepsilon$, $\Delta_\gamma$
and $P_\beta$ in Eq.~\eqref{contact} by $-i \sigma^i_{\ \varepsilon}$,
$-i \sigma^n_{\ \varepsilon}$, $-i \sigma^n_{\ \gamma}$ and $-i \sigma^n_{\ \beta}$ or
contract $-i \sigma^n_{\ \alpha}$ with Eq.~\eqref{determinationnogamma}, therefore there
are 5 equalities

\begin{gather}
    \det\begin{vmatrix}
     g^{\alpha \alpha'} & P^{\alpha'} & \Delta^{\alpha'} & n^{\alpha'} & g^{i
     \alpha'}\\
     P^{\alpha} & P^2 & 0 & P \cdot n & \frac{1}{2} \Delta^i\\
     \Delta^{\alpha} & 0 & \Delta^2 & \Delta \cdot n & \mathscr{P}^i\\
     n^{\alpha} & P \cdot n & \Delta \cdot n & 0 & - i \sigma^{i n}\\
     g^{\alpha j} & P^j & \Delta^j & 0 & 0
   \end{vmatrix} \doteq 0, \quad
   \det\begin{vmatrix}
     g^{\alpha \alpha'} & P^{\alpha'} & \Delta^{\alpha'} & n^{\alpha'} &
     n^{\alpha'}\\
     P^{\alpha} & P^2 & 0 & P \cdot n & \frac{1}{2} \Delta \cdot n\\
     \Delta^{\alpha} & 0 & \Delta^2 & \Delta \cdot n & \mathscr{P} \cdot n\\
     n^{\alpha} & P \cdot n & \Delta \cdot n & 0 & 0\\
     g^{\alpha j} & P^j & \Delta^j & 0 & i \sigma^{j n}
   \end{vmatrix} \doteq 0, \notag \\
    \det\begin{vmatrix}
     g^{\alpha \alpha'} & P^{\alpha'} & n^{\alpha'} & n^{\alpha'} & g^{i
     \alpha'}\\
     P^{\alpha} & P^2 & \frac{1}{2} \Delta \cdot n & P \cdot n & P^i\\
     \Delta^{\alpha} & 0 & \mathscr{P} \cdot n & \Delta \cdot n
     & \Delta^i\\
     n^{\alpha} & P \cdot n & 0 & 0 & 0\\
     g^{\alpha j} & P^j & i \sigma^{j n} & 0 & 0
   \end{vmatrix} \doteq 0, \quad
    \det\begin{vmatrix}
     g^{\alpha \alpha'} & n^{\alpha'} & \Delta^{\alpha'} & n^{\alpha'} & g^{i
     \alpha'}\\
     P^{\alpha} & \frac{1}{2} \Delta \cdot n & 0 & P \cdot n & P^i\\
     \Delta^{\alpha} & \mathscr{P} \cdot n & \Delta^2 & \Delta
     \cdot n & \Delta^i\\
     n^{\alpha} & 0 & \Delta \cdot n & 0 & 0\\
     g^{\alpha j} & i \sigma^{j n} & \Delta^j & 0 & 0
   \end{vmatrix} \doteq 0,\label{determinationsigma1}\\
   \det\begin{vmatrix}
     n^{\alpha'} & P^{\alpha'} & \Delta^{\alpha'} & n^{\alpha'} & g^{i
     \alpha'}\\
     \frac{1}{2} \Delta \cdot n & P^2 & 0 & P \cdot n & P^i\\
     \mathscr{P} \cdot n & 0 & \Delta^2 & \Delta \cdot n &
     \Delta^i\\
     0 & P \cdot n & \Delta \cdot n & 0 & 0\\
     i \sigma^{j n} & P^j & \Delta^j & 0 & 0
   \end{vmatrix} \doteq 0,\notag
\end{gather}
where we have used some known on-shell identities~\cite{Cotogno:2019vjb,Fu:2022rkn,Fu:2022bpf}
and $\sigma^{i j}$ is removed due to the symmetry from the definition~\eqref{gluontransversitygpds}. Analogously, respectively replacing $g^j_\tau$, $g^j_\tau$,
$\Delta_\rho$ and $P_\nu$ in Eq.~\eqref{contact} by $-i \sigma^j_{\ \tau}$, $-i \sigma^n_{\ \tau}$,
$-i \sigma^n_{\ \rho}$ and $-i \sigma^n_{\ \nu}$ or
contracting $-i \sigma^n_{\ \alpha'}$ with Eq.~\eqref{determinationnogamma} gives
\begin{gather}
    \det\begin{vmatrix}
     g^{\alpha \alpha'} & P^{\alpha'} & \Delta^{\alpha'} & n^{\alpha'} & g^{i
     \alpha'}\\
     P^{\alpha} & P^2 & 0 & P \cdot n & P^i\\
     \Delta^{\alpha} & 0 & \Delta^2 & \Delta \cdot n & \Delta^i\\
     n^{\alpha} & P \cdot n & \Delta \cdot n & 0 & 0\\
     - g^{j \alpha} & \frac{1}{2} \Delta^j & \mathscr{P}^j & - i
     \sigma^{j n} & 0
   \end{vmatrix} \doteq 0, \quad
   \det\begin{vmatrix}
     g^{\alpha \alpha'} & P^{\alpha'} & \Delta^{\alpha'} & n^{\alpha'} & g^{i
     \alpha'}\\
     P^{\alpha} & P^2 & 0 & P \cdot n & P^i\\
     \Delta^{\alpha} & 0 & \Delta^2 & \Delta \cdot n & \Delta^i\\
     n^{\alpha} & P \cdot n & \Delta \cdot n & 0 & 0\\
     - n^{\alpha} & \frac{1}{2} \Delta \cdot n & \mathscr{P} \cdot n & 0 & i \sigma^{i n}
   \end{vmatrix} \doteq 0, \notag \\
   \det\begin{vmatrix}
     g^{\alpha \alpha'} & P^{\alpha'} & \Delta^{\alpha'} & n^{\alpha'} & g^{i
     \alpha'}\\
     P^{\alpha} & P^2 & 0 & P \cdot n & P^i\\
     - n^{\alpha} & \frac{1}{2} \Delta \cdot n & \mathscr{P} \cdot n & 0 & i \sigma^{i n}\\
     n^{\alpha} & P \cdot n & \Delta \cdot n & 0 & 0\\
     g^{\alpha j} & P^j & \Delta^j & 0 & 0
   \end{vmatrix} \doteq 0, \quad
   \det\begin{vmatrix}
     g^{\alpha \alpha'} & P^{\alpha'} & \Delta^{\alpha'} & n^{\alpha'} & g^{i
     \alpha'}\\
     - n^{\alpha} & \frac{1}{2} \Delta \cdot n & \mathscr{P} \cdot n & 0 & i \sigma^{i n}\\
     \Delta^{\alpha} & 0 & \Delta^2 & \Delta \cdot n & \Delta^i\\
     n^{\alpha} & P \cdot n & \Delta \cdot n & 0 & 0\\
     g^{\alpha j} & P^j & \Delta^j & 0 & 0
   \end{vmatrix} \doteq 0, \label{determinationsigma2} \\
   \det\begin{vmatrix}
     - n^{\alpha} & \frac{1}{2} \Delta \cdot n & \mathscr{P} \cdot n & 0 & i \sigma^{i n}\\
     P^{\alpha} & P^2 & 0 & P \cdot n & P^i\\
     \Delta^{\alpha} & 0 & \Delta^2 & \Delta \cdot n & \Delta^i\\
     n^{\alpha} & P \cdot n & \Delta \cdot n & 0 & 0\\
     g^{\alpha j} & P^j & \Delta^j & 0 & 0
   \end{vmatrix} \doteq 0.\notag
\end{gather}

Solving Eqs.~\eqref{determinationsigma1} and~\eqref{determinationsigma2} can
get the corresponding on-shell identities to eliminate $i \sigma^{n i} \otimes $($n^{[ \alpha' \nobracket}
g^{\nobracket \alpha ] j }$, $ \Delta \cdot n \, n^{\{ \alpha' \nobracket}
g^{\nobracket \alpha \} j} -2 n^{\alpha'} n^\alpha \Delta^j $, $ P \cdot n \,
P^{[ \alpha' \nobracket} g^{\nobracket \alpha ] j}- P^{[ \alpha' \nobracket}
n^{\nobracket \alpha ]} P^j$ and $\Delta \cdot n \, P^{\{ \alpha' \nobracket}
g^{\nobracket \alpha \} j}- P^{\{ \alpha' \nobracket} n^{\nobracket \alpha \}} \Delta^j$) terms.

Taking the changes, $g^i_\varepsilon \rightarrow \gamma_\varepsilon$,
$n_\delta \rightarrow \gamma_\delta$ or $g^j_\tau \rightarrow \gamma_\tau$,
$n_\sigma \rightarrow \gamma_\sigma$, from Eq.~\eqref{contact} gives
\begin{equation}\label{determinationgamma1}
    \det\begin{vmatrix}
     g^{\alpha \alpha'} & P^{\alpha'} & \Delta^{\alpha'} & n^{\alpha'} & 0\\
     P^{\alpha} & P^2 & 0 & P \cdot n & M\\
     \Delta^{\alpha} & 0 & \Delta^2 & \Delta \cdot n & 0\\
     n^{\alpha} & P \cdot n & \Delta \cdot n & 0 & \slashed{n}\\
     g^{\alpha j} & P^j & \Delta^j & 0 & \gamma^j
   \end{vmatrix} \doteq 0, \quad
   \det\begin{vmatrix}
     g^{\alpha \alpha'} & P^{\alpha'} & \Delta^{\alpha'} & 0 & g^{i \alpha'}\\
     P^{\alpha} & P^2 & 0 & M & P^i\\
     \Delta^{\alpha} & 0 & \Delta^2 & 0 & \Delta^i\\
     n^{\alpha} & P \cdot n & \Delta \cdot n & \slashed{n} & 0\\
     g^{\alpha j} & P^j & \Delta^j & \gamma^j & 0
   \end{vmatrix} \doteq 0,
\end{equation}
and
\begin{equation}\label{determinationgamma2}
    \det\begin{vmatrix}
     g^{\alpha \alpha'} & P^{\alpha'} & \Delta^{\alpha'} & n^{\alpha'} & g^{i
     \alpha'}\\
     P^{\alpha} & P^2 & 0 & P \cdot n & P^i\\
     \Delta^{\alpha} & 0 & \Delta^2 & \Delta \cdot n & \Delta^i\\
     n^{\alpha} & P \cdot n & \Delta \cdot n & 0 & 0\\
     0 & M & 0 & \slashed{n} & \gamma^i
   \end{vmatrix} \doteq 0,\quad
   \det\begin{vmatrix}
     g^{\alpha \alpha'} & P^{\alpha'} & \Delta^{\alpha'} & n^{\alpha'} & g^{i
     \alpha'}\\
     P^{\alpha} & P^2 & 0 & P \cdot n & P^i\\
     \Delta^{\alpha} & 0 & \Delta^2 & \Delta \cdot n & \Delta^i\\
     0 & M & 0 & \slashed{n} & \gamma^i\\
     g^{\alpha j} & P^j & \Delta^j & 0 & 0
   \end{vmatrix} \doteq 0.
\end{equation}
Combining Eqs.~\eqref{determinationgamma1} and \eqref{determinationgamma2}, one
can obtain that the tensor structures $\left( \Delta^i + 2 \xi P^i \right)
\left( \Delta \cdot n \gamma^j -\slashed{n} \Delta^j \right) P^{\alpha'} P^\alpha$ and
$\left( \Delta^i + 2 \xi P^i \right)\left( P \cdot n \gamma^j -\slashed{n} P^j \right)
g^{\alpha' \alpha}$ can be represented by others.

\bibliographystyle{unsrt}
\bibliography{ref}

\end{document}